\documentclass{egpubl}
\usepackage{sca2020}
\usepackage{amssymb}
\usepackage{multirow}
\usepackage{amsmath}
\usepackage[ruled,linesnumbered]{algorithm2e}
\usepackage{kotex}
\usepackage{xcolor}
\newcommand{\argmax}{\arg\!\max} 
\JournalPaper         
\usepackage[T1]{fontenc}
\usepackage{dfadobe}  

\usepackage{cite}  
\BibtexOrBiblatex
\electronicVersion
\PrintedOrElectronic
\ifpdf \usepackage[pdftex]{graphicx} \pdfcompresslevel=9
\else \usepackage[dvips]{graphicx} \fi

\usepackage{egweblnk} 

\title{Functionality-Driven Musculature Retargeting}

\author[Ryu et al.]
{\parbox{\textwidth}{\centering Hoseok Ryu$^{1}$, Minseok Kim$^{1}$, Seungwhan Lee$^{1}$, Moon Seok Park$^{2}$, Kyoungmin Lee$^{2}$, Jehee Lee$^{1}$ 
        }
        \\
{\parbox{\textwidth}{\centering $^1$Seoul National University, Korea\\
         $^2$Seoul National University Bundang Hospital, Korea
      }
}
}

\begin{document}

\teaser{
 \includegraphics[width=.95\linewidth]{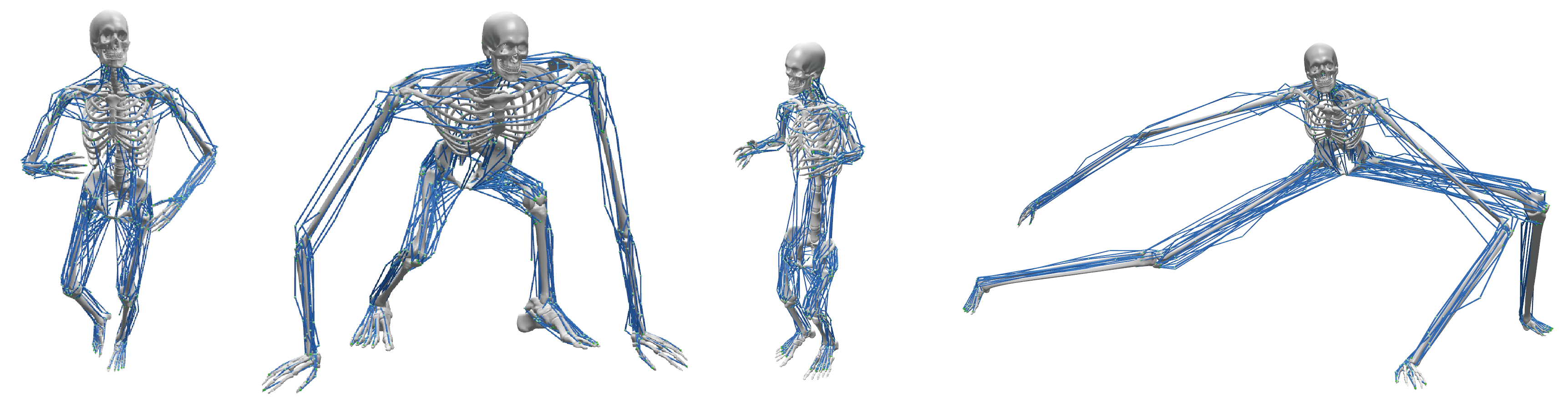}
 \centering
  \caption{\label{fig:teaser} Various musculoskeletal models of different body sizes and proportions. The full-body musculature of the reference model on the left was retargeted to the others. (Left to right) Musculoskeletal model : reference model, Hulk, Dwarf, and Alien.}
\label{fig:teaser}
}

\maketitle
\begin{abstract}
We present a novel retargeting algorithm that transfers the musculature of a reference anatomical model to new bodies with different sizes, body proportions, muscle capability, and joint range of motion while preserving the functionality of the original musculature as closely as possible. The geometric configuration and physiological parameters of musculotendon units are estimated and optimized to adapt to new bodies. The range of motion around joints is estimated from a motion capture dataset and edited further for individual models. The retargeted model is simulation-ready, so we can physically simulate muscle-actuated motor skills with the model. Our system is capable of generating a wide variety of anatomical bodies that can be simulated to walk, run, jump and dance while maintaining balance under gravity. We will also demonstrate the construction of individualized musculoskeletal models from bi-planar X-ray images and medical examinations.



\end{abstract}  
\section{Introduction}

The human body in computer graphics has evolved from a stick figure with torque actuators to a muscle-driven skeleton. The musculoskeletal model for physics-based simulation includes many parts and parameters that require careful orchestration to make it anatomically plausible. Designing such a musculoskeletal model is labor-intensive and hand-crafted models are often inaccurate to actuate an anatomical body in physically simulated environments. We present a new method that retargets a reference musculoskeletal model to new bodies of different sizes, body proportions, muscle capability, and joint Range of Motion (ROM). 

The musculature is a system of muscles and tendons actuating the skeleton. The motion of the skeleton is the result of the harmonious coordination of many muscles. Muscles are correlated with each other in various ways.  A group of muscles works together to actuate a single joint and the ROM around the joint is also determined by multiple adjacent muscles. Some adjacent muscles share a tendon and thus the contraction of one muscle may affect the activation of the others. A single (multi-articular) muscle may actuate two or more joints simultaneously.

The functional role of each individual skeletal muscle is thoroughly studied and well documented in anatomy textbooks~\cite{gray2009gray}. Each muscle is responsible for particular movement (e.g., flexion/extension, adduction/abduction, internal/external rotation) of body parts. For example, the active contraction (shortening) of the rectus femoris (a muscle in the thigh) results in knee extension and hip flexion. Even if the muscle is inactive, its background elasticity generates passive force to prevent the adjacent joints from excessive movement, which influences the ROM around the joints.

Our focus is on adapting the musculature of a reference anatomical model to new bodies while preserving the functionality of the musculature. The muscles are expected to perform the same functional roles and have the same ROMs in the new body after retargeting. The key parameters that affect muscle functions and joint ROMs are the length of musculotendons and their geometric routing paths. The core of our retargeting algorithm is a numerical solver that optimizes the key parameters throughout the entire body. The algorithm allows us to create a variety of musculoskeletal models including exotic creatures.

We also implemented a human body editing system that allows the user to interactively edit the height, width, bone length, bending and torsion angles, and joint ROMs of the skeleton. The musculature of the anatomical model is automatically retargeted to fit the new skeletal body. Additionally, the user is allowed to edit physics parameters, such as the mass and inertia of individual body parts, and make the body stronger or weaker by adjusting Hill-type muscle parameters, such as force-length curves and force-velocity curves. The physics and muscle parameters are exploited in a physically based simulation of muscle-actuated characters.

The simulation and control of under-actuated biped locomotion under gravity have been studied for decades in computer graphics and robotics. The recent progress in deep reinforcement learning research has been very successful in continuous control problems and made it possible to simulate dynamic human activities without compromising fundamental physics laws~\cite{peng2018deepmimic,lee2019scalable}. We will demonstrate that our muscle-actuated anatomical model is compatible with state-of-the-art simulation algorithms. Our retargeted characters with extreme body proportions can be physically simulated and controlled to walk, run, kick, and jump while maintaining balance under gravity.

\section{Related Work}

There has been a stream of studies for reproducing natural human motion by incorporating the increasingly more accurate models of human anatomy and the mechanics of anatomical structures. Muscle-based anatomical modeling and simulation have been extensively explored and exploited in Computer Graphics and Biomechanics.

\subsubsection*{Muscle Modeling}
The anatomical model includes an articulated skeleton driven by its musculature. The Hill-type muscle model~\cite{hill1938heat,zajac1989muscle,millard2013flexing} has been broadly adopted to encode the nonlinear contraction dynamics of muscles. The geometry of a muscle is often simplified into a sequence of line segments for the efficiency of computation. There have been continuous efforts in Computer Animation and Biomechanics to demonstrate the computational plausibility of volumetric FEM muscles~\cite{lee2009comprehensive,si2014realistic,fan2014active,berranen2014real}. {\em In vivo} estimation of muscle parameters relies primarily on medical 3D imaging. Matias et al.~\shortcite{matias2009transformation} studied the estimation of muscle-bone attachments based on bony landmarks. Levin et al.~\shortcite{levin2011extracting} studied the estimation of muscle fiber directions from diffusion tensor images. Arnold et al.~\shortcite{arnold2010model} and Holzbaur et al.~\shortcite{holzbaur2005model} provided a comprehensive reference of muscle modeling parameters for lower and upper limbs. Anatomical modeling for specific body parts, such as face~\cite{sifakis2005automatic,ichim2017phace}, feet~\cite{park2018multi}, hands~\cite{Sueda2008Musculotendon,sachdeva2015biomechanical}, shoulders~\cite{van1994finite,kaptein2004estimating,maurel2000human}, tongue~\cite{stavness2012automatic}, and jaw~\cite{zoss2018empirical}, has been explored for last two decades. Comprehensive full-body musculoskeletal modeling and simulation systems are also available freely~\cite{Delp2007opensim} and commercially~\cite{Damsgaard2006anybody}. Muscle routing plays an important role in muscle-actuated full-body simulation. Many modeling approaches, such as conditional waypoints~\cite{delp1990interactive}, obstacle sets~\cite{garner2000obstacle}, surface SSD~\cite{murai2016anatomographic}, and wrapping surfaces~\cite{rajagopal2016full}, have been exploited to improve the estimate of muscle lengths during joint motion.

\subsubsection*{Joint Modeling}
The joints of a skeletal model are often simplified as either revolute (1 degree of freedom) or ball-and-socket (3 degrees of freedom) joints, though anatomical joints are structurally more complicated than simple mechanical joints~\cite{lee2008spline}. Maurel and Thalman~\shortcite{maurel2000human} designed a skeletal model of human shoulders and represented the flexibility of the ball-and-socket joint by a combination of the sinus cone and an interval of the arm twist angle. Lee~\shortcite{lee2000hierarchical} studied a general approach of describing the range of motion around a joint based on quaternion half-spaces and their boolean combination. Akhter and Black~\shortcite{akhter2015pose} collected a motion capture dataset that explores a wide range of human poses, and learned the pose-dependent range of motion around each joint. Jiang and Liu~\shortcite{jiang2018data} represented the boundary of valid human joint configurations by using a fully-connected neural network. Measuring the joint range of motion and muscle lengths are of great clinical interest. A variety of clinical examination techniques are performed frequently in clinical practice~\cite{reese2016joint}.

\subsubsection*{Muscle-Driven Simulation and Control}
Recent accomplishments in Computer Animation made it possible to reproduce realistic human motion in physics-based simulation~\cite{liu2012terrain,coros2010generalized,lee2010data,sok2007simulating}. The emergence of deep reinforcement learning accelerated the progress and successfully simulated skillful actions, such as jump, flip, cartwheel, basketball dribbling, and even aerobatic flapping flight~\cite{peng2018deepmimic,liu2018learning,yu2018learning,won2018aerobatics,lee2019scalable}. Muscle-driven anatomical simulation poses further challenges of dealing with the complexity of anatomical modeling and scalability in physics-based simulation. Wang et al.~\shortcite{wang2012optimizing} demonstrated a muscle-actuated biped with eight musculotendon units on each leg. The biped was essentially two-dimensional because all muscles are aligned in the sagittal plane. Geijtenbeek et al.~\shortcite{geijtenbeek2013flexible} improved the flexibility in biped design by allowing off-sagittal muscles and optimizing muscle-attachments sites for locomotion control. Comprehensive 3D musculoskeletal models with up to 120 musculotendon units were successfully simulated and controlled using a two-level control architecture that consists of low-level muscle coordination based on quadratic programming and higher-level gait modulation based on stochastic optimization~\cite{lee2014locomotion}. Lee et al.~\shortcite{lee2018dexterous} demonstrated that volumetric FEM muscles are viable for dextrous manipulation, such as juggling, that requires a high level of control precision. Nakada et al.~\shortcite{Nakada2018biomimetic} built a comprehensive full-body neuromuscular system to learn biomimetic sensorimotor control of human animation. Seth et al. ~\cite{seth2018opensim} 
built an open-source software {\em OpenSim} which provides extensible neuromusculoskeletal models of humans and animals. The system allows the creation, physical simulation, and biomechanical analysis of neuromusculoskeletal models. The muscle-actuated control system proposed by Lee et al.~\shortcite{lee2019scalable} employed a hierarchical deep reinforcement learning algorithm to cope with the full details (46 joints and 346 muscles) of the musculoskeletal model and successfully simulated dynamic motor skills. They also demonstrated pre-operative and post-operative simulation of pathologic gaits. We built our musculoskeletal model on top of their simulation system, which provides favorable features such as analytic differentiation of muscle-actuated dynamic systems, muscle routing based on LBS (linear blend skinning), and efficient handling of joint ROMs based on LCP (linear complementary problem). LCP provides an effective way of dealing with inequality conditions which prevent muscle over-extension in the simulation of muscle contraction dynamics.

\begin{figure}
\begin{center}
    \includegraphics[width=0.38\textwidth]{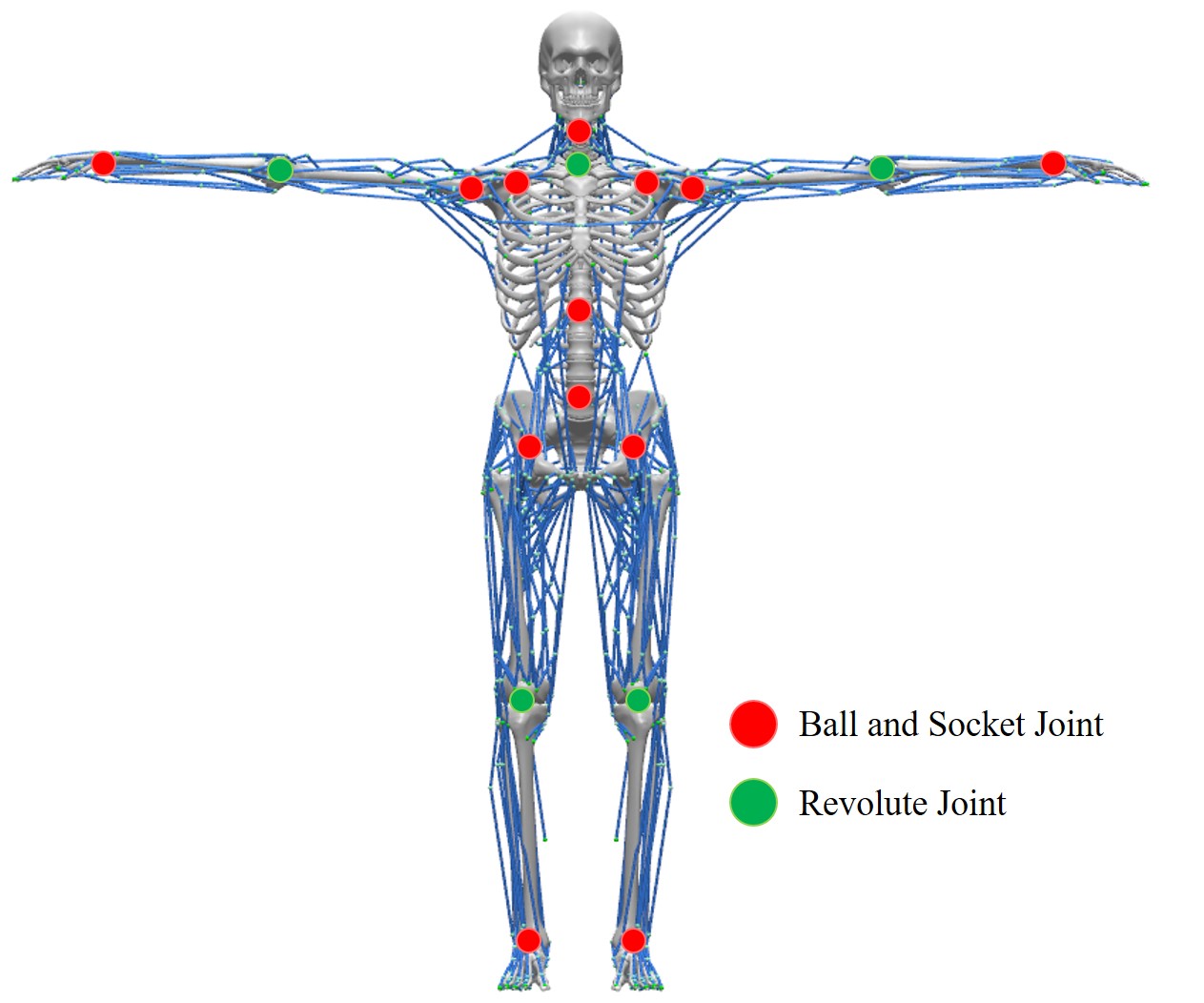}
    \caption{Our reference musculoskeletal model. }
    \label{fig:reference_model}
\end{center}
\end{figure}

\subsubsection*{Anatomy Transfer and Retargeting}
Constructing a high-fidelity anatomical model is time-consuming and labor-intensive. Therefore, there has been a series of studies that retarget a geometry of reference anatomical model to a range of target bodies. Dicko et al.~\shortcite{ali2013anatomy} transferred the geometry of internal anatomy from an input template to a target body while maintaining anatomic constraints. Saito et al.~\shortcite{saito2015computational} generated a spectrum of human body shapes with various degrees of muscle
growth. Kadle{\v{c}}ek et al.~\shortcite{kadlevcek2016reconstructing} generated a personalized anatomic model by retargeting a template model to fit full-body 3D scans while accounting for body shape variations and pose variations. Our work is on the line of these approaches with emphasis that our retargeted model is optimized to function as closely as possible to its reference model in the realm of muscle-driven simulation and control.

\section{Musculoskeletal System}

Our musculoskeletal model has an articulated skeleton with 5 revolute joints, 13 ball-and-socket joints, and 6 DoFs (Degrees of Freedom) at the skeletal root yielding 50 DoFs in total (see Figure~\ref{fig:reference_model}). Our work is focusing on simulating human motion driven by four limbs and spinal joints. 
The model includes 282 musculotendon units corresponding to skeletal muscles, which are attached to bones on each end by tendons. The attachment site on the proximal end is called the {\em origin} of the muscle, while the attachment site on the opposite end is called its {\em insertion}. Muscle contraction pulls the attached bones and moves the joint between them. Our model comprehensively includes skeletal muscles that contribute to the motion in any limb joints. Since our model has no fingers or toes, the muscles that originate and insert within hands or feet are omitted.

\begin{figure}
\begin{center}
    \includegraphics[width=0.20\textwidth]{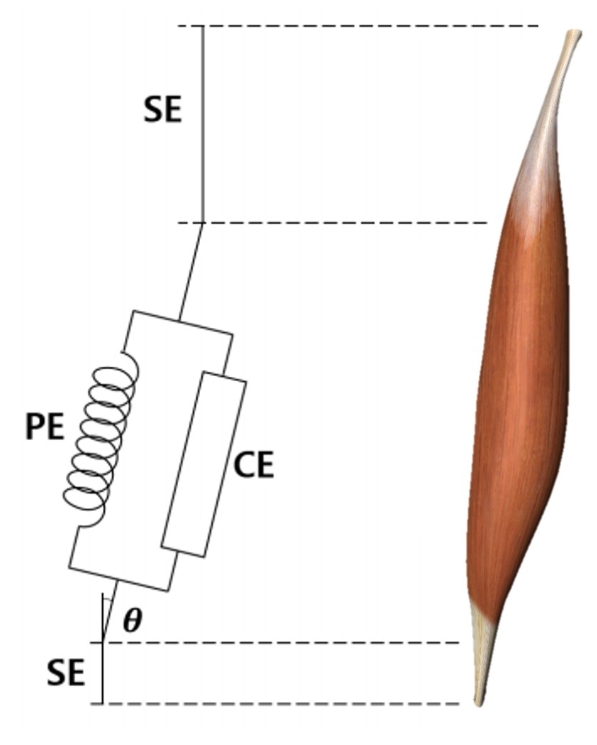}
    \includegraphics[width=0.25\textwidth]{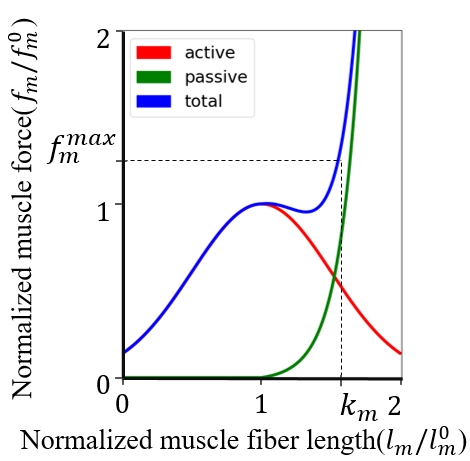} 
    \caption{Hill-type model of musculotendon units. The force-length curve plots maximum isometric force of the active fiber and passive fiber force as a function of normalized muscle fiber length. }
    \label{fig:Hill_type_muscle}
\end{center}
\end{figure}

\subsection{Muscle Model}

The Hill-type muscle is a three-element scheme that models the muscle contraction-force relation (see Figure~\ref{fig:Hill_type_muscle}). A musculotendon unit is defined by an active contractile element (CE), a passive parallel element (PE), and a passive serial element (SE). The active force of the contractile element comes from muscular contraction. All skeletal muscles have a rest length. When the muscle is stretched to its ideal length, it can maximize muscular contraction. The serial element represents the tendons on either side of the contractile segment of the muscle. The tendon is a tough band of fibrous connective tissue that connects muscle to bone. The parallel element models the background elasticity of the muscle. The parallel element is responsible for muscle passive behavior when it is stretched. 

Let $l_\textrm{m}$, $l_\textrm{t}$ and $l_\textrm{mt}$ be muscle fiber length, tendon length, and the total muscle-tendon length, respectively. Optimal fiber length $l_\textrm{m}^0$ is the length of muscle fibers when it develops maximum isometric force. Let $l_\textrm{t}^0$ be tendon slack length. Pennation $\theta$ is the angle between the muscle fibers and the tendon axis. Assuming a quasi-static setting, the tension in muscle fibers is a function of muscle length and its activation $a$, while the tension in the tendon is a function of only muscle length. The tension in the three elements holds
\begin{equation}
\label{eq:muscle_tension}
    F = \big(f_\mathrm{CE}(\tilde{l}_\textrm{m},a) + f_\mathrm{PE}(\tilde{l}_\textrm{m})\big)\cdot\cos{\theta} = f_\mathrm{SE}(\tilde{l}_\textrm{t}),
\end{equation}
where $\tilde{l}_\textrm{m}=l_\textrm{m}/l_\textrm{m}^0$ and $\tilde{l}_\textrm{t}=l_\textrm{t}/l_\textrm{t}^0$ are normalized muscle and tendon lengths, respectively. Here, we assume that the pennation angle is constant regardless of muscle contraction.



\begin{figure}
    \centering
    \includegraphics[width=0.45\textwidth]{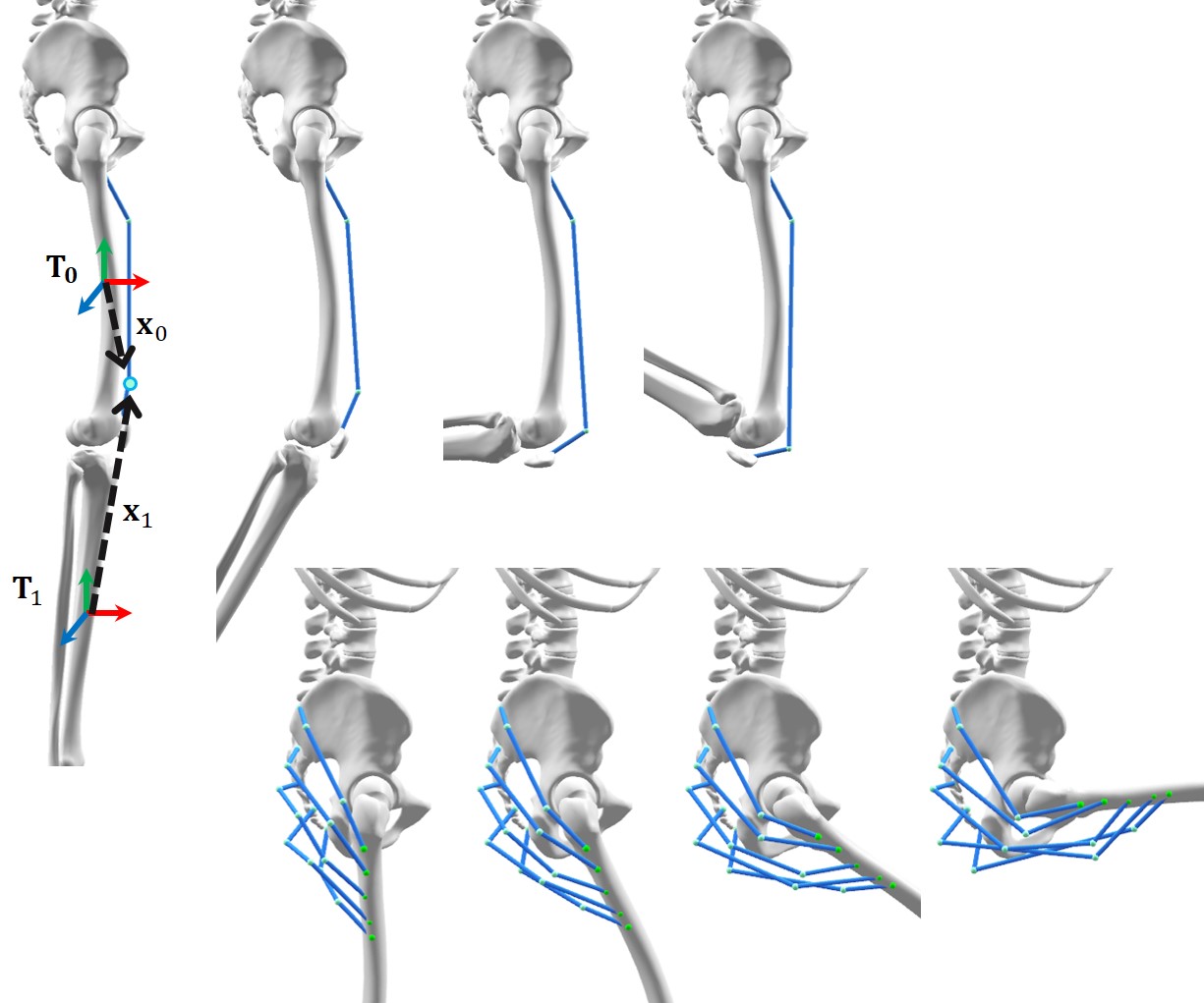}
    \caption{Polyline approximation of (Top) Vastus Intermedius and (Bottom) Gluteus Maximus. Since the waypoints are represented in LBS coordinates, they move during knee/hip flexion to better fit muscle deformation. }
    \label{fig:waypoints}
\end{figure}

Actual muscles wrap around bones and soft tissues during joint motion and they are also constrained by surrounding ligaments and skin tissue layers. Therefore, the straight line segment between the origin and insertion sites of a muscle is a lousy approximator of its actual length. Better approximators can be achieved by exploiting geometric proxies, such as waypoints~\cite{delp1990interactive}, obstacle sets~\cite{garner2000obstacle}, and wrapping surfaces~\cite{rajagopal2016full}, that approximate muscle deformation during joint motion. We use waypoints as geometric proxies and expressed the location of a waypoint relative to nearby bones using the idea of linear blend skinning (LBS) to improve the accuracy and flexibility over bone-attached waypoints (see Figure~\ref{fig:waypoints}).
 \begin{equation}
     \mathbf{p} = \sum_j w_{j}\mathbf{T}_{j}\mathbf{x}_j
 \end{equation}
where $\mathbf{T}_j\in\mathbb{R}^{4\times4}$ is the transformation matrix of $j$-th bone, $w_{j}$ is skinning weight, and $\mathbf{x}_j\in\mathbb{R}^3$ is the coordinates relative to $j$-th bone. Since the derivatives of LBS-based waypoints can be derived in an analytic form, they are well-suited for efficient dynamics simulation with inequality constraints~\cite{lee2019scalable}.

\subsection{Parametric Skeleton Model}

\begin{figure}
\begin{center}
    \includegraphics[width=0.45\textwidth]{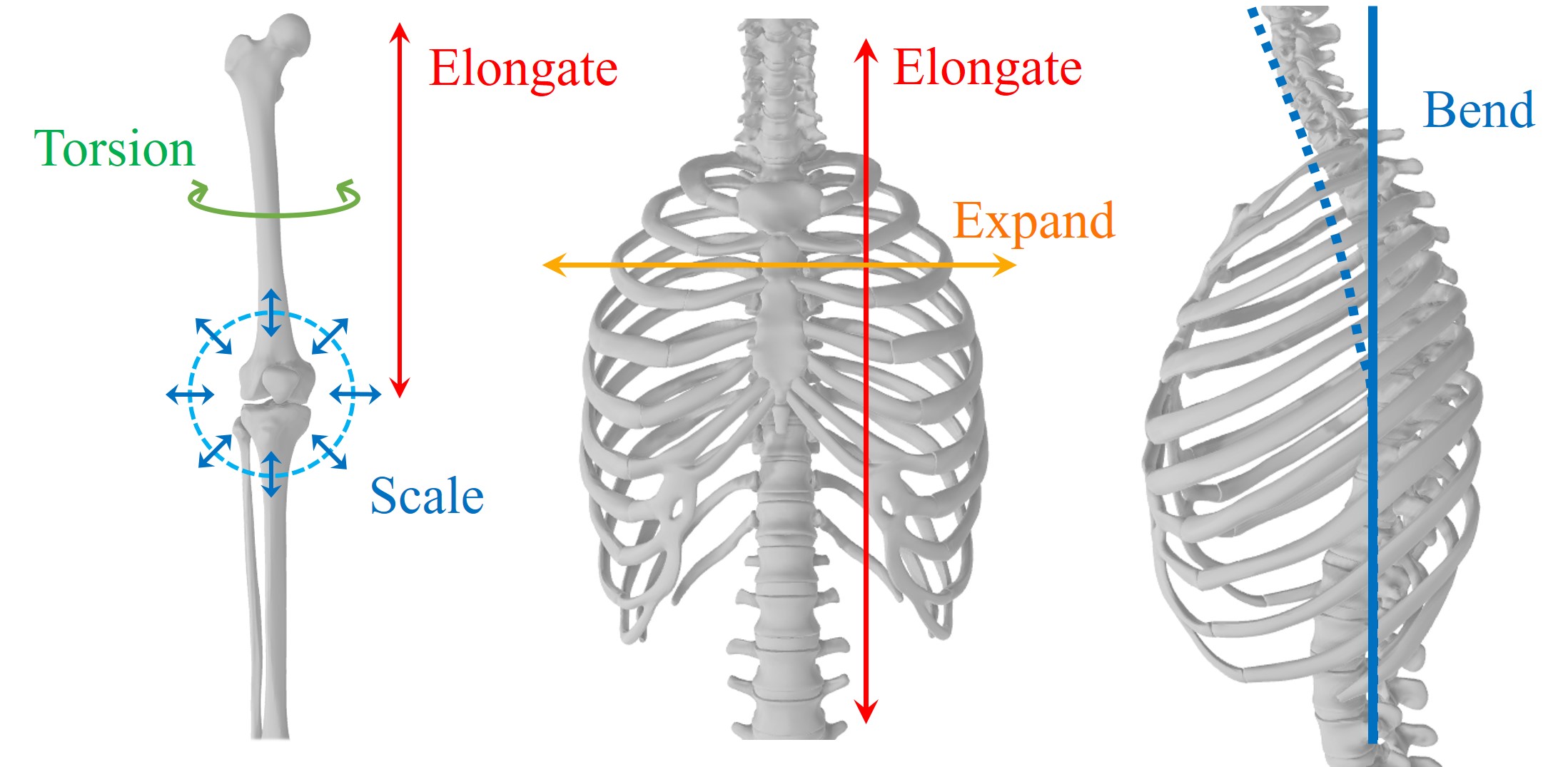}
    \caption{ Parametric modeling of the trunk and limbs. }
    \label{fig:Parametric_Skeleton}
\end{center}
\end{figure}

The skeletal body is a parametric model that can generate a wide variety of human body shapes (see Figure~\ref{fig:Parametric_Skeleton}). The model has a reference shape and a set of shape parameters. The trunk and four limbs are parameterized to have variations in their length, size, and alignment relative to the reference shape. The limb bones including Femur, Tibia, Humerus, Ulna exhibit an elongated shape with its proximal head, long shaft, and distal head. The limb bone has four parameters: proximal head scale, distal head scale, shaft elongation, and torsion. Even though the trunk consists of many bones, the trunk is considered as a lumped body and parameterized by three parameters: elongation, expand, and bend. The hands and feet have one parameter for scaling their size. We use a geometric deformation method \cite{kadlevcek2016reconstructing} to produce parametrically-varied shapes from the reference model.

We scaled physics and muscle parameters relative to geometric parameters inspired by the work of Hodgins and Pollard~\shortcite{hodgins1997adapting}. Assuming uniform scaling by a factor $L$ in all dimensions, they suggested to scale time, force, mass, moment of inertia, velocity, stiffness, and damping by factors $L^{1/2}$, $L^3$, $L^3$, $L^5$, $L^{1/2}$, $L^2$, and $L^{5/2}$, respectively. This scaling rule provides a rough guideline as to how physics and muscle parameters should be adjusted relative to body scaling. The user can edit the mass and inertia of each individual body part and scale the force-length curve of each muscle following the guideline. The guideline is not strict since stable dynamics simulation can be often accomplished for a wide range of parameters. 

\section{Joint Range of Motion Modeling}

Many articulated characters in computer graphics have the ROM around joints described by an interval of min/max angles for each DoF. This approach is inherently limited because the pose-dependence of joint ROMs cannot be captured in per-DoF intervals. An alternative approach is to estimate a high-dimensional function from a motion capture dataset, which fits the boundary of valid poses. The {\em isValid} function $\mathcal{V}(\mathbf{q}):\mathbb{R}^N\rightarrow\{0,1\}$ takes a full-body pose as input and outputs a boolean value to answer whether the pose is valid or not. The function estimation requires nonlinear regression over a large collection of full-body poses~\cite{akhter2015pose,jiang2018data}. The musculoskeletal system readily provides an {\em isValid} function without requiring nonlinear regression because the musculoskeleton represents the fundamental mechanism of joint movements and their limits.

\subsection{Muscle-Induced ROM}

The joint ROM is affected by many anatomic factors, such as bony structures, surrounding ligaments, joint capsules, and muscular tension. We are particularly interested in the factors induced by muscles. The elasticity of muscle fibers represented by a parallel element determines the distance a joint can move to its full potential in the direction of the muscle. This extension of muscle stretching is described by an implicit function $C_i(\mathbf{q}):\mathbb{R}^{N}\rightarrow\mathbb{R}$, where $\mathbf{q}$ is the generalized coordinates of the full-body skeletal pose, and $N$ is its dimension. The constraint induced by passive fiber tension is
\begin{equation}
    C_i(\mathbf{q}) = k_\mathrm{m} l_\mathrm{m}^0 - l_\mathrm{m}(\mathbf{q}) ~\ge 0,
\label{eq:passive_constraint}
\end{equation}
where muscle length $l_\mathrm{m}(\mathbf{q})$ is a function of skeletal pose and $k_\mathrm{m}$ is the maximum ratio of extension that the muscle fiber can stretch. Note that there exists a direct mapping between skeletal poses and muscle lengths. Given a pose, the lengths of all muscles are uniquely determined and vice versa. We assume that passive elongation reaches its limit when muscle tension is larger than maximum isometric tension in muscle fibers by a certain margin (see Figure~\ref{fig:Hill_type_muscle}(right)). 
Deriving from muscle contraction dynamics~\cite{thelen2003adjustment}, we set the maximal muscle-tendon length by
\begin{equation}
    l_\mathrm{mt}^\mathrm{max} = k_\mathrm{m} l_\mathrm{m}^0 + k_\mathrm{t} l_\mathrm{t}^0,
\label{eq:max_length}
\end{equation}
where $k_\mathrm{m}=1.6$ and $k_\mathrm{t}=1.03$. 
We can also express other factors induced by bones and ligaments as implicit functions (e.g., the constraint by the kneecap can be denoted by a function of skeletal pose). The collection of all constraints defines an {\em isValid} function.
\begin{equation}
    \mathcal{V}(\mathbf{q}) = \left\{\begin{array}{ll}
        1, & \mathrm{if~~~} C_i(\mathbf{q}) \ge 0 \mathrm{~~~for~~~} \forall i \\
        0, & \mathrm{otherwise} \\
    \end{array} \right.
\end{equation}

It is often the case that multiple muscles cross a joint and all those muscles affect the ROM in the joint. Among those muscles, some are multi-articular meaning that it crosses multiple joints. The presence of multi-articular muscles makes joint ROM pose-dependent. For example, we have a wider range of motion in the hip when the knee is flexed than it is fully extended. Our {\em isValid} function based on the musculoskeletal model inherently captures the pose-dependence of ROM.

\begin{figure}
    \centering
    \includegraphics[width=0.22\textwidth]{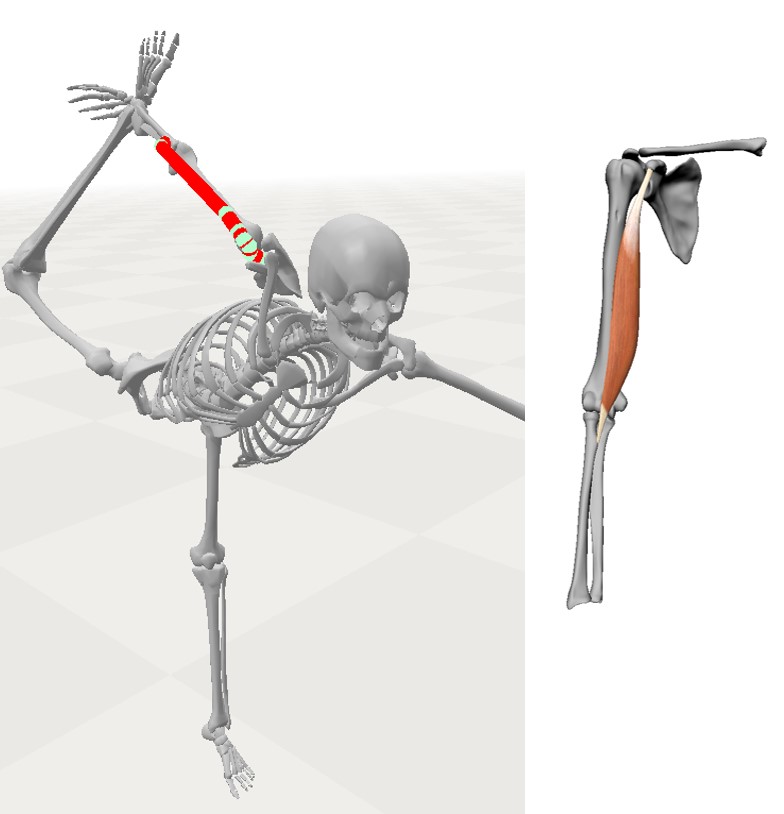}
    \includegraphics[width=0.22\textwidth]{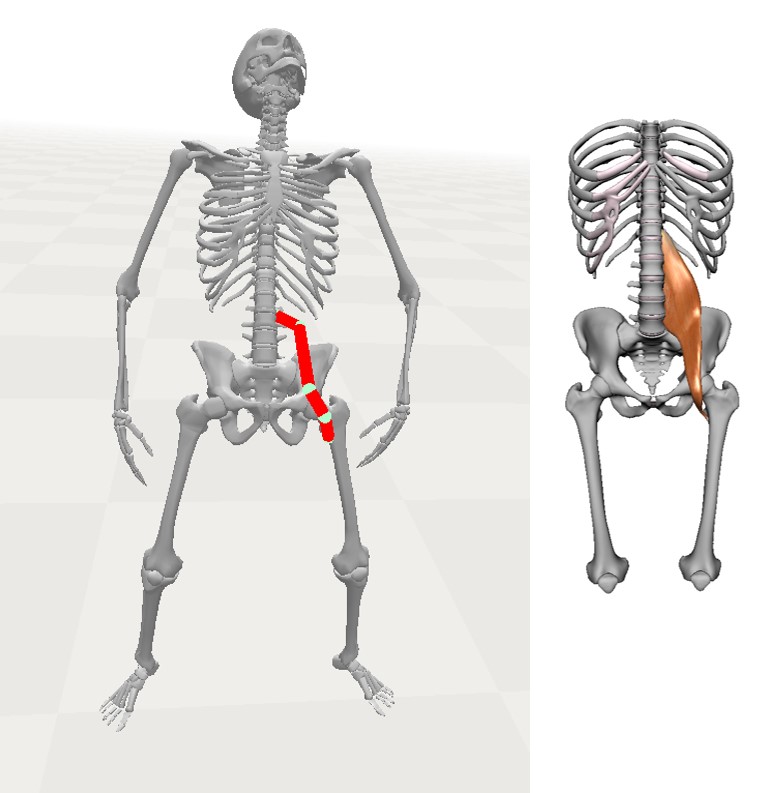}
    \caption{Full-body poses that maximize individual muscle lengths. (Left) Right Bicep Brachii Short Head. (Right) Left Psoas Major.}
    \label{fig:maximal_stretch}
\end{figure}

\subsection{Muscle Length Estimation}
\label{sec:length_estimation}

Even though carefully specifying muscle parameters would yield appropriate joint ROMs, parameter tuning is laborious and time-consuming. We estimate muscle parameters (in particular, muscle fiber and tendon lengths) automatically such that the {\em isValid} function fits the boundary of full-body poses captured in a motion dataset. Specifically, we used a data set collected by Akhter and Black~\shortcite{akhter2015pose}, which includes an extensive variety of stretching and yoga poses performed by trained athletes and gymnasts. In the dataset, we can find a full-body pose that maximizes the length $l_\mathrm{mt}$ for each muscle (see Figure~\ref{fig:maximal_stretch}). Assuming that muscle fiber and tendon length ratios are invariant, solving Equation~(\ref{eq:muscle_tension}) and Equation~(\ref{eq:max_length}) gives us optimal fiber length $l_\mathrm{m}^0$ and tendon slack length $l_\mathrm{t}^0$ at the maximal length $l_\mathrm{mt}^\mathrm{max}$. This estimation of muscle lengths guarantees that all poses in the dataset are valid.

The dataset actually provides us with the lower bound for the maximal length of individual muscles and therefore it tends to under-estimate the maximal lengths. For example, the range of knee extension is bounded by the bone and ligament structures around the knee rather than muscle passive force. It is highly likely that knee flexor muscles (e.g., Hamstring and Gastrocnemius) do not extend to their maximal length when the knee is straight. Hence, the dataset does not provide a precise estimation of the maximal length of the knee flexors. We address this under-estimation issue by measuring joint torques at several key-poses, for which muscles should not develop passive force. The key-poses include standing, zero-gravity, and T-poses. We gradually increase $l_\mathrm{m}^0$ and $l_\mathrm{t}^0$ of muscles as they develop passive force until the force magnitude is reduced below a certain threshold.

\subsection{ROM Editing}
\label{sec:rom_editing}

Muscle lengths thus estimated yield an {\em isValid} function $\mathcal{V}(\mathbf{q})$, which serves as a reference ROM model. The reference model represents the level of flexibility of athletes, which is far higher than the flexibility level of average, non-athletic individuals. We would like to be able to edit or modify the reference model to create various individualized ROM functions. ROM editing involves improving or reducing the ROM in a particular joint and shifting the range. Note that ROM editing is a supplementary step that provides flexible and convenient user control over the modeling and retargeting procedures. The ROM editing step can be skipped if it is not necessary.

Since $\mathcal{V}(\mathbf{q})$ is a high-dimensional function defined implicitly depending on many factors and complex anatomical structures, it is not practical to edit the function directly. Alternatively, we define a new, modified function $\mathcal{\tilde V}(\mathbf{q})=\mathcal{V}(T(\mathbf{q}))$ indirectly with transformation $T$ over the input pose. The action of $T$ affects the ROM inversely. For example, if the transformation exaggerates poses, the ROM with the modified function $\mathcal{\tilde V}$ becomes narrower than the reference model because the test pose will be exaggerated first and then passed to the reference {\em isValid} test $\mathcal{V}$. 

\begin{figure}
    \centering
    \includegraphics[width=\columnwidth]{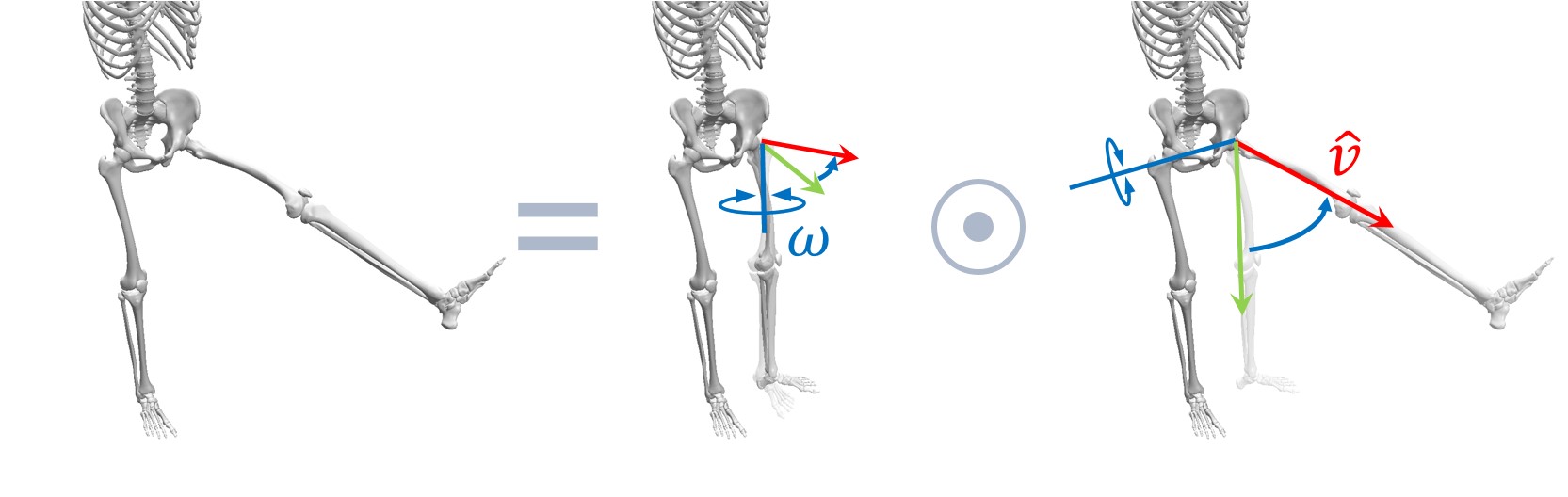}
    \caption{3D Rotation Decomposition into 1D axial rotation followed by 2D conic rotation. The axis of the second rotation is orthogonal to the first axis. The second rotation is conic since it can move the bone shaft within a cone.}
    \label{fig:rotation}
\end{figure}

We can define the transformation for each joint or each DoF independently. For a revolute joint, the ROM is simply an interval $[\phi-\psi, \phi+\psi]$, where $\phi$ is the average of the joint angles in the dataset. The transformation $T(\theta) = s(\theta-\phi)+\phi+t$ of joint angle $\theta$ in pose $\mathbf{q}$ results in $s$-scaling of the interval followed by $t$-shifting. Note that we do not need to know the value of $\psi$ to define the transformation. Similarly, we can define ROM-editing transformations for ball-and-socket joints. The configuration of a ball-and-socket joint is a three-dimensional rotation, which can be decomposed into a twist about the principle (bone shaft) axis in the reference pose followed by another rotation about its orthogonal axis (see Figure~\ref{fig:rotation}). The decomposed configuration is $(\omega,\hat{v})$, where $\omega\in\mathbb{R}$ is a twist angle and $\hat v\in\mathbb{S}^2$ is a unit vector indicating the direction of the bone shaft. The ROM of $\omega$ is a simple interval, while the ROM of $\hat v$ is a cone with its center direction estimated from the dataset. The scaling and shifting transformation on the cone is also defined in a straightforward manner and can be exploited for ROM editing. Through simple user interfaces, the user can edit ROMs interactively by modulating scale and shift factors.

\section{Musculature Retargeting}

\begin{figure*}
\begin{center}
    \includegraphics[width=\textwidth]{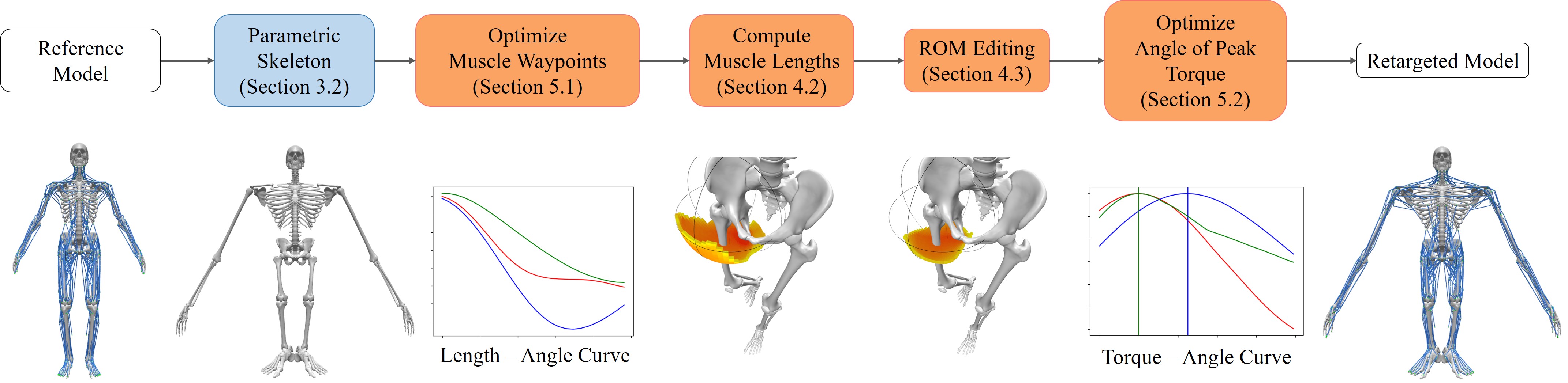}
    \caption{ The overview of musculature retargeting. }
    \label{fig:overview}
\end{center}
\end{figure*}

In this section, we describe a retargeting algorithm that adapts the full-body musculature to the change of the skeleton (see Figure~\ref{fig:overview}). Our parametric skeletal model can generate a rich variety of human body shapes. The goal of retargeting is to make the varied models viable for physics-based simulation and control. 

\begin{figure}
\begin{center}
    \includegraphics[width=\columnwidth]{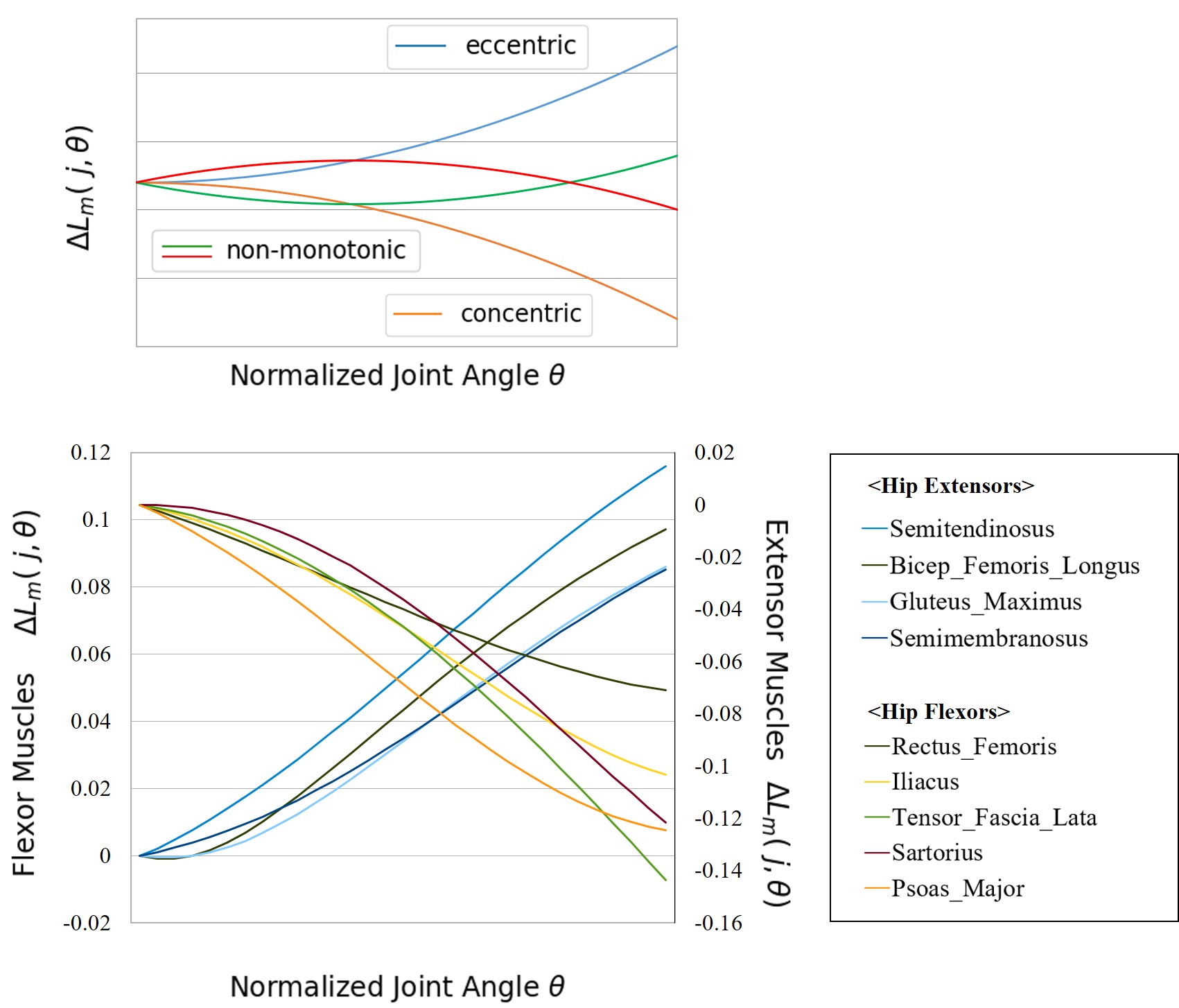}
    \caption{ Types of length-angle curves. (Top) The monotonically decreasing (or increasing) curve indicate that the muscle is an agonist (or antagonist) during the joint motion represented by angle $\theta$.  The Non-monotonic curve indicates that the muscle is maybe functionally insignificant for the particular joint motion. (Bottom) Examples of hip flexors and hip extensors. }
    \label{fig:length_angle_curve}
\end{center}
\end{figure}

The functional roles of skeletal muscles are classified by the plot of their length changes and the direction of muscle forces acting on the attached bones. For example, a muscle is called a flexor (or extensor) of the joint if the muscle shortens as the joint flexes (or extends). Table~\ref{tab:muscle_activity} shows examples of muscle actions during various joint motions. The muscle is functionally significant for causing a particular joint motion if its length-angle curve is increasing (as an agonist) or decreasing (as an antagonist)~(see Figure~\ref{fig:length_angle_curve}). Therefore, the length-angle curves characterize the functional roles of muscles. Na\"ive scaling of the musculoskeletal model could alter the monotonicity and modality of the length-angle curves, muscle force directions at attachment sites and joint ROMs, eventually affecting the functionality of muscles in the new body.


\begin{table}[t]
    \caption{Examples of muscle activities during joint motion. The full version with details can be found in anatomy textbooks~\cite{gray2009gray}. *Biarticula r (two-joint) muscles.}
    \label{tab:muscle_activity}
    \small{
    \begin{tabular}{|l|l|}
        \hline \textbf{Joint Motion} & \textbf{Active Muscles} \\ \hline
        Hip Flexor & Psoas major, Iliacus, Rectus femoris*, \\
                   & Tensor fasciae latae*, Sartorius* \\ \hline

Hip Extensor & Gluteus maximus, Biceps femoris*, \\
             & Semitendinosus*, Semimembranosus*  \\ \hline

Hip  & Gluteus medius, Pectineus, Tensor fasciae latae*\\ 
Medial Rotator & \\ \hline
Hip             & Quadratus femoris, Obtrator externus, \\
Lateral Rotator & Biceps femoris*  \\ \hline

Hip Abductor & Gluteus minimus \& medius, \\
             & Tensor fasciae latae*\\ \hline

Hip Adductor & Adductor longus \& brevis \& magnus \\ \hline

Knee Extensor & Vastus medialis \& intermedius \& laterlias, \\
              & Rectus femoris*  \\ \hline

Knee Flexor & Sartorius*, Biceps femoris*, Semitendinosus*, \\
            & Semimembranosus*, Gastrocnemius*, Plantaris* \\ \hline

Foot          & Gastrocnemius*, Plantaris*, Soleus, \\
Plantarflexor & Tibialis posterior\\ \hline
                    
Foot          & Tibialis anterior, Extensor hallucis longus, \\
Dorsiflexor   & Extensor digitorum longus, Fibularis tertius\\ \hline
    \end{tabular}
    }
\end{table}

Musculature retargeting involves three categories of parameters. The parameters in each category have the desired impacts on muscle functionality. Our algorithm solves for three categories of parameters sequentially. 
\begin{itemize}
    \item Optimize muscle routing to modulate length-angle curves and force directions, where the muscle routing is expressed by {\em LBS coordinates of muscle waypoints}. 
    \item Determine {\em maximal muscle lengths} to achieve desired ROMs (as described in Section~\ref{sec:length_estimation} and Section~\ref{sec:rom_editing}). 
    \item Optimize {\em muscle fiber and tendon length ratios} to modulate the angle of peak torque at joints.
\end{itemize}
Since we already discussed the second step in the previous section, this section focuses on the first and third steps which are formulated as nonlinear optimization.


\subsection{Muscle Routing Optimization}

Waypoints play an important role in estimating muscle lengths during joint motion. While the use of LBS coordinates improves the estimation accuracy substantially, manual tuning of LBS coordinates is laborious. Our algorithm decides LBS coordinates of waypoints such that the energy function is minimized:
\begin{equation}
    E_\mathrm{waypoint} = E_\mathrm{direction} + w_l E_\mathrm{length}.
\end{equation}
The first term encourages muscle force directions through the polyline approximations of muscles to be preserved
\begin{equation}
    E_\mathrm{direction}(m) = \sum_k \sum_j \int_{\theta=0}^1 
         \|f_m(k; j,\theta) \times f'_m(k; j,\theta) \|^2  d\theta,
\end{equation}
where $m$ is the index of muscles, $k$ is the index of waypoints of the muscle $m$ (including the origin and insertion), $j$ is the type of joint motion in which the muscle $m$ participates, and $\theta\in[0,1]$ is the normalized joint angle. $f$ and $f'$ are normalized force directions in the reference and retargeted bodies, respectively, at point $k$ to $k+1$. $f$ and $f'$ are determined by LBS coordinates.

\begin{figure}
\begin{center}
    \includegraphics[width=0.45\textwidth]{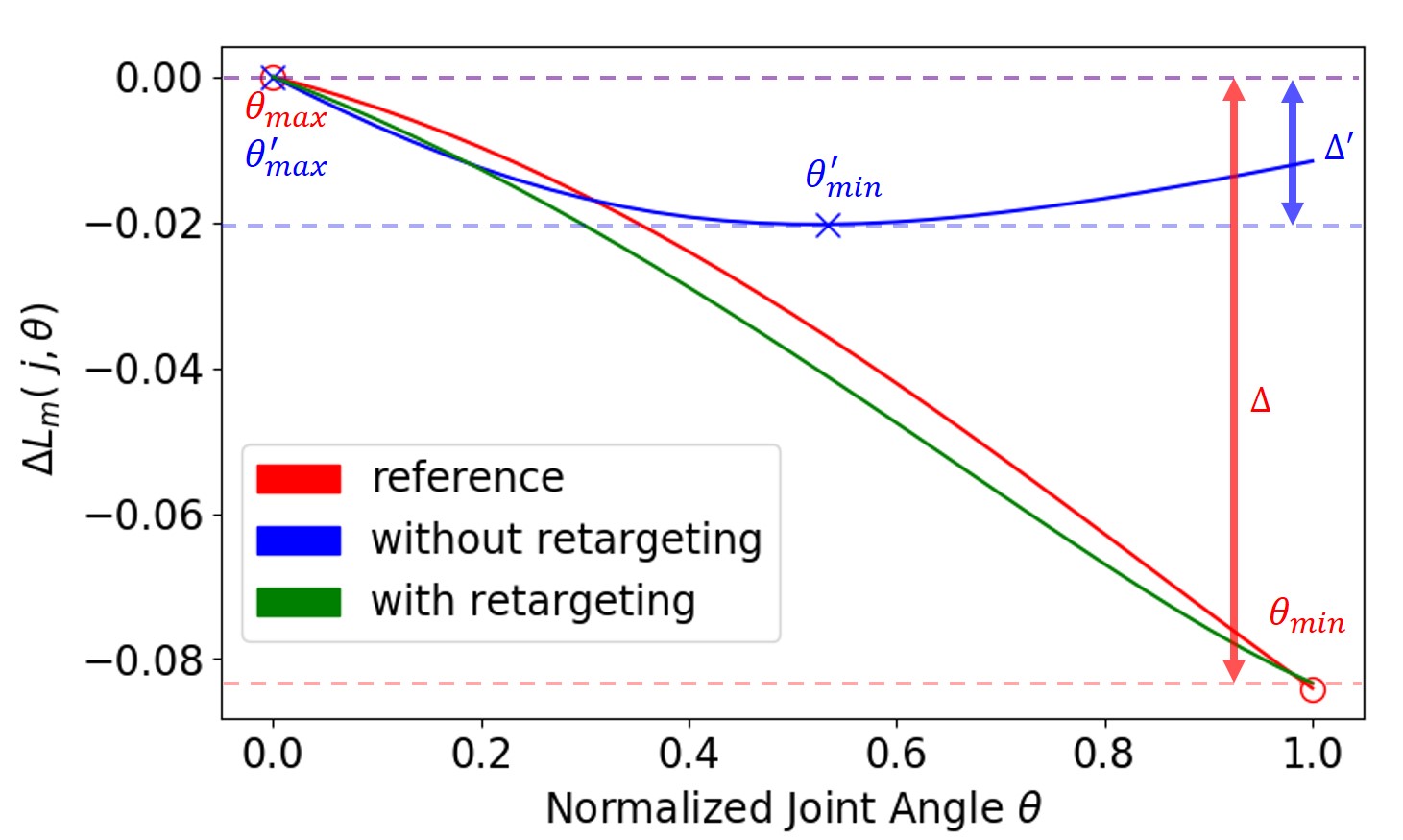}
    \caption{ Muscle retargeting. (Red) The length-angle curve of Biceps Brachii Short Head during elbow flexion. (Blue) $60\%$ shortening of Humerus and Ulna/Radius yields a length-angle curve that is non-monotonic. (Green) Muscle retargeting makes the curve closer to its original shape. }
    \label{fig:muscle_retarget}
\end{center}
\end{figure}

The second term encourages the qualitative characteristics of length-angle curves to be preserved in the retargeted body. Assuming that the length-angle curve $L_{m}(j, \theta)$ is monotonic or unimodal on interval $[0,1]$, the curve can be characterized by three parameters: $\theta_\mathrm{max}$, $\theta_\mathrm{min}$, $\Delta$. The curve has its maximum value at $\theta_\mathrm{max}$, minimum value at $\theta_\mathrm{min}$, and the difference between the maximum and minimum value is $\Delta$ (see Figure~\ref{fig:muscle_retarget}). The second term is defined using the characteristics parameters.
\begin{equation}
    E_\mathrm{length}(m) = \sum_j (\theta_\mathrm{max}-\theta'_\mathrm{max})^2 +
    (\theta_\mathrm{min}-\theta'_\mathrm{min})^2 +
    w_\Delta(\Delta - \Delta')^2.
\end{equation}
We use a gradient descent algorithm to minimize the energy function. The gradient of the energy function is estimated by finite difference and line search along gradient direction accelerates the algorithm. In our example, $w_l=10$ and  $w_\Delta=50$.

The performance and quality of gradient-based optimization depend on the initial guess of the optimized parameters. During the scaling of the skeleton, each waypoint moves together with the skeleton anchored to the nearest point on the bone. The anchor position relative to the skeleton serves as an initial guess of the optimization.


\subsection{Angle of Peak Torque}

Joint torque is the sum of muscle torques crossing the joint. Peak torque usually occurs in the joint ROM when the muscle fibers are close to their optimal length. For example, knee and elbow joints generate their maximum bending torque when they are moderately flexed. The angle of peak torque is related to muscle fiber and tendon length ratios. 

Peak torque angle adjustment is a constrained optimization process based on gradient descent. The optimization objective is that all joints in the retargeted model have their peak torque at the same angles as in the reference model. The optimization parameters are fiber-tendon length ratios at all muscles. The ratios are constrained to vary within $30\%$ from their original values.
\begin{equation}
    E_\mathrm{apt} = \sum_j ||\theta_\mathrm{ref}(j) -  \underset{\theta}{\argmax}\; {\tau(j,\theta)}||^2,
\end{equation}
where $\theta_\mathrm{ref}$ is the angle of peak torque in the reference model and $\tau(j,\theta)$ is the net torque in joint motion $j$ when its normalized angle is $\theta$. 

\section{Experimental Results}

We can generate a wide spectrum of body proportions through interactive user interfaces. The user interface provides a set of slide bars to manipulate body parameters and allows the user to edit joint ROMs. In this section, we show experiments conducted with four representative characters: {\em Human}, {\em Hulk}, {\em Dwarf}, and {\em Alien} (see Figure~\ref{fig:teaser} and Table~\ref{tab:PPer}). The Human is a reference model we created based on a human skeleton geometry. We annotated origins, insertions, and waypoints of all muscles and tuned their parameters to create a model that can be used for physics-based simulation. The Hulk has a large muscular body with long arms and short legs and the trunk leans slightly forward. The Alien has a short torso and extremely long limbs, while the Dwarf is the opposite with a long torso and short limbs. The musculature of the Human is retargeted to the others to build musculoskeletal dynamic systems.

\begin{table}
\caption{Skeleton scaling parameters. Positive angle denotes clock-wise torsion. Left and Right sides are symmetrically adjusted.} 
\centering 
\small
\begin{tabular}{l c c c c c}
\hline
  & & Reference & Alien & Hulk & Dwarf \\
\hline \hline
Trunk & Elongate &1.0 & 1.0 & 1.0 & 2.0\\
      & Expand & 1.0 & 1.0 & 2.0 & 0.6\\
      & Mass   & 1.0 & 1.0 & 1.0 & 2.0\\

\hline
Femur & Elongate &1.0 & 2.5 & 0.8 & 0.6\\
      & Torsion & $0.0^{\circ}$ & $0.0^{\circ}$ & $10.0^{\circ}$ & $-10.0^{\circ}$ \\
      & Mass & 1.0 & 2.5 & 0.8 & 0.6\\
\hline
Tibia & Elongate &1.0 & 2.5 & 0.8 & 0.6\\
      & Torsion & $0.0^{\circ}$ & $0.0^{\circ}$ & $10.0^{\circ}$ & $10.0^{\circ}$ \\
      & Mass & 1.0 & 2.5 & 0.8 & 0.6\\

\hline

Humerus & Elongate &1.0 & 2.5 & 1.8 & 0.6\\
        & Torsion & $0.0^{\circ}$ & $-10.0^{\circ}$ & $10.0^{\circ}$ & $0.0^{\circ}$ \\
        & Mass & 1.0 & 2.5 & 1.8 & 0.6\\

\hline
Ulna    & Elongate &1.0 & 2.5 & 1.4 & 0.6\\
        & Torsion & $0.0^{\circ}$ & $10.0^{\circ}$ & $0.0^{\circ}$ & $0.0^{\circ}$ \\
        & Mass   & 1.0 & 2.5 & 1.4 & 0.6\\

\hline
Neck    & Elongate &1.0 & 2.0 & 1.0 & 1.0 \\
        & Mass     & 1.0 & 2.0 & 1.0 & 1.0 \\

\hline 
\end{tabular}
\label{tab:PPer}
\end{table}


\subsection{ROM Construction}

\begin{figure}
\begin{center}
    \includegraphics[width=\columnwidth]{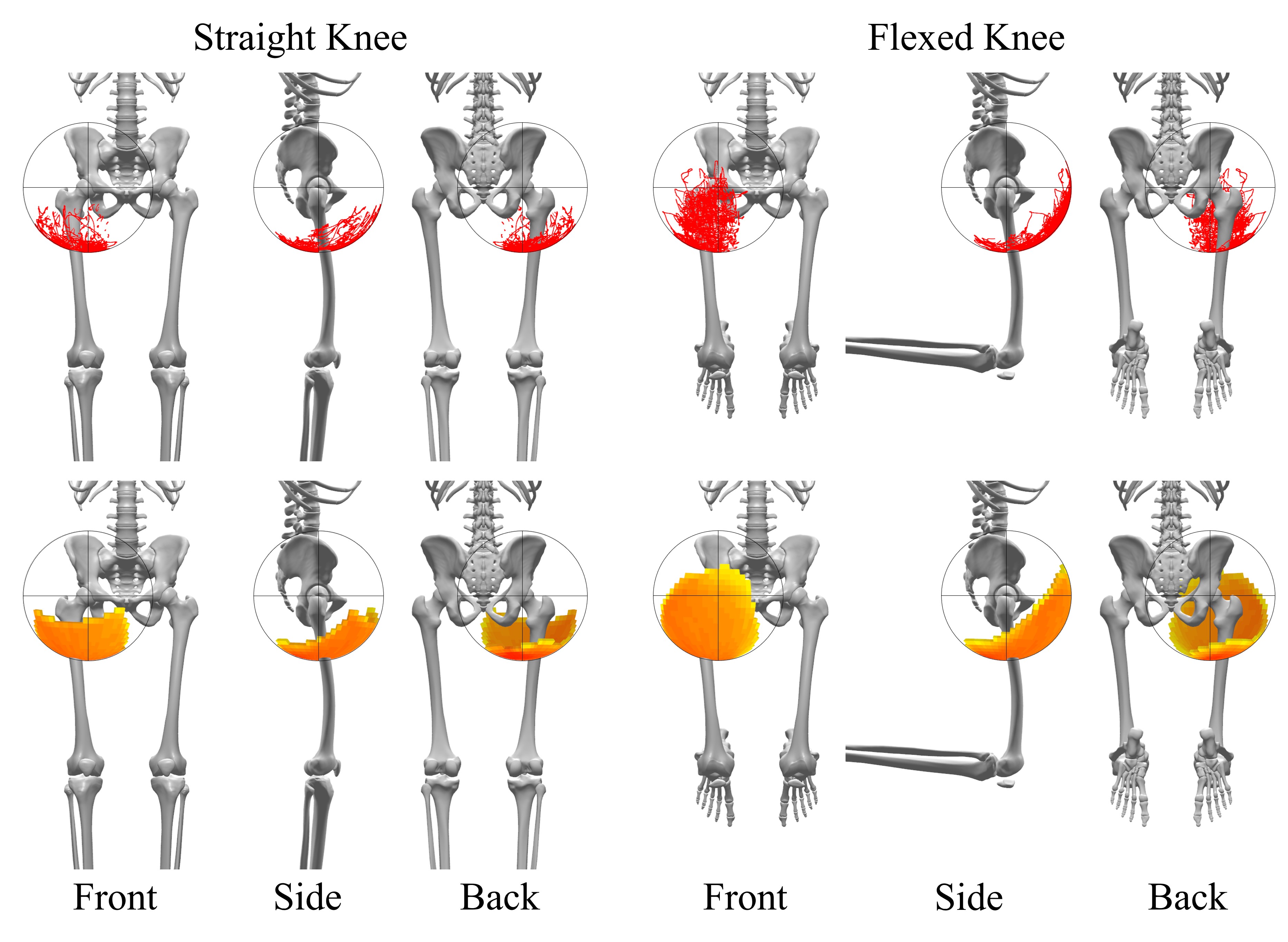}
    \caption{ The pose-dependent ROM of the hip joint. (Top) Valid hip configurations captured in the reference dataset.
    The range on the left figures (the knee angles smaller than 30-degree) is narrower than the range on the right figures (the knee angles larger than 90-degree).
    (Bottom) The muscle-induced hip ROM of our musculoskeletal model. The hip configuration is decomposed into a twist angle and the direction of the bone shaft. The bone shaft direction is depicted as a yellow-red range on the unit sphere at the hip. The color spectrum indicates the range of the twist angle. The yellow color represents narrow ranges, while the red color represents wider ranges.}
    \label{fig:ROM_comparison}
\end{center}
\end{figure}

We constructed the ROMs of our reference model from a motion capture data set~\cite{akhter2015pose}. We excluded corrupted outliers from the dataset, doubled the data by mirror reflection, and subsampled at a 1/10 ratio to take approximately 70,000 full-body poses. The mirror reflection makes the ROMs symmetric. The maximum lengths of all muscles are estimated from the pose set and further relaxed to address the under-estimation issue as explained in Section~\ref{sec:length_estimation}. Muscle lengths thus estimated represent pose-dependent joint limits.
Figure~\ref{fig:ROM_comparison} shows that the hip ROM is wider when the knee is flexed than it is straight. This observation is due to the presence of biarticular muscles. The hamstring muscles (Biceps femoris, Semitendinosus, and Semimembranosus) are all biarticular, originating at the bottom of the Pelvis, crossing both the hip and knee joints, and inserting at the head of the Tibia. The hamstring muscles are therefore involved in both hip extension and knee flexion. The hamstring muscles are extended when the knee is straight and thus the potential for the hip flexion is limited.

\begin{figure}
\begin{center}
    \includegraphics[width=1.0\columnwidth]{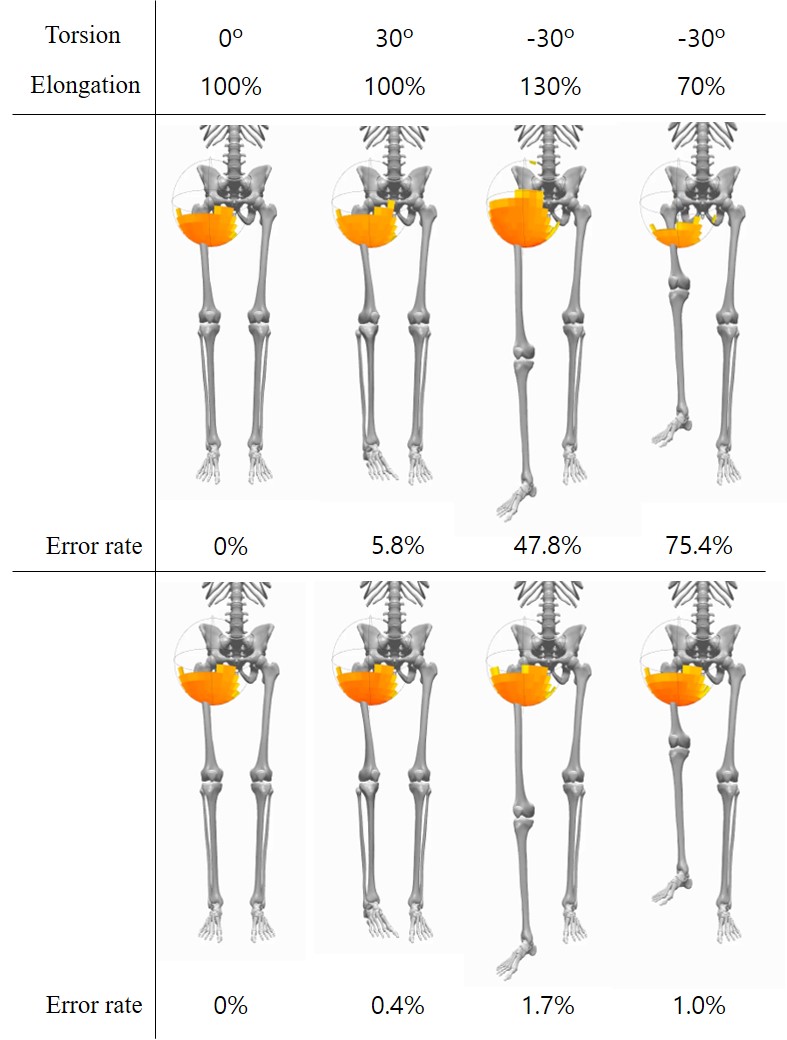}
    \caption{ The hip ROM affected by the elongation and torsion of the femur. (TOP) Na{\"i}ve linear scaling of the muscles results in undesirable changes in the ROM. (Bottom) Our retargeting algorithm precisely recalculates the muscle lengths to preserve the ROM of the reference model regardless of body shape variations. 
    }
    \label{fig:mocap_validation}
\end{center}
\end{figure}

The ROM around a joint is strongly influenced by the change of the body shape. Figure~\ref{fig:mocap_validation} shows how the longitudinal elongation (in the range of 70\% to 130\% of the original length) and torsion (in the range of -30$^\circ$ to 30$^\circ$) of the femur affect the hip ROM. Na{\"i}ve linear scaling of the muscles attached to the femur results in undesirable changes in the hip ROM, which is discretized into $18\times36\times36$ cells. Each cell has a boolean value. We measured the error rate by the percentage of false positives and true negatives in the cells. Without retargeting, the error rates are in the range of 5.8\% to 75.4\%. The maximum error drops down to 1.7\% after retargeting.



\begin{figure}
\begin{center}
    \includegraphics[width=\columnwidth]{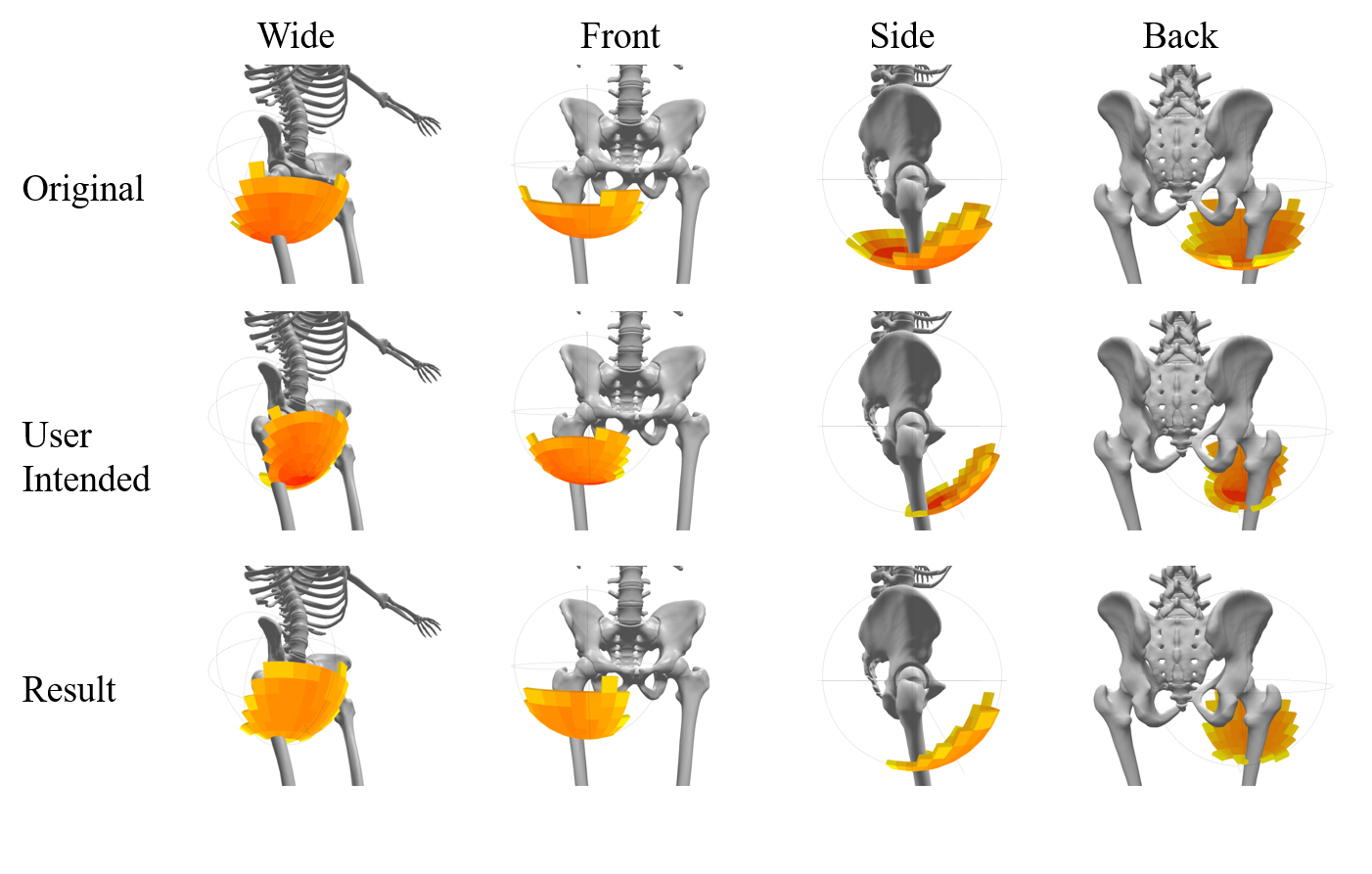}
    \caption{ Interactive ROM editing. (Top) The estimated ROM of the hip joint. (Middle) The user edits the ROM interactively by defining transformations that make the ROM shift forward and narrower in its width. (Bottom) Recalculating muscle lengths match the muscle-induced ROM to the user's intention. }
    \label{fig:rom_editing}
\end{center}
\end{figure}

Our system provides a simple user interface that allows the user to edit the ROM of any joint interactively. The user can make the ROM wider, narrower, or shifted on the space of conic and twist rotations. The editing operations are defined by transformations as explained in Section~\ref{sec:rom_editing}. Our system recalculates the muscles at relevant sites such that the muscle-induced ROM matches the user's intention approximately. In Figure~\ref{fig:rom_editing}, the hip ROM tilted forward by 30$^\circ$ and shrank in its width on the cone by a factor of 0.63 by user-specified transformations. Recalculating muscle lengths approximates the user editing with error rate of 7.3\%.

\subsection{Muscle Routing Evaluation}

\begin{figure*}
\begin{center}
    \includegraphics[width=.88\textwidth]{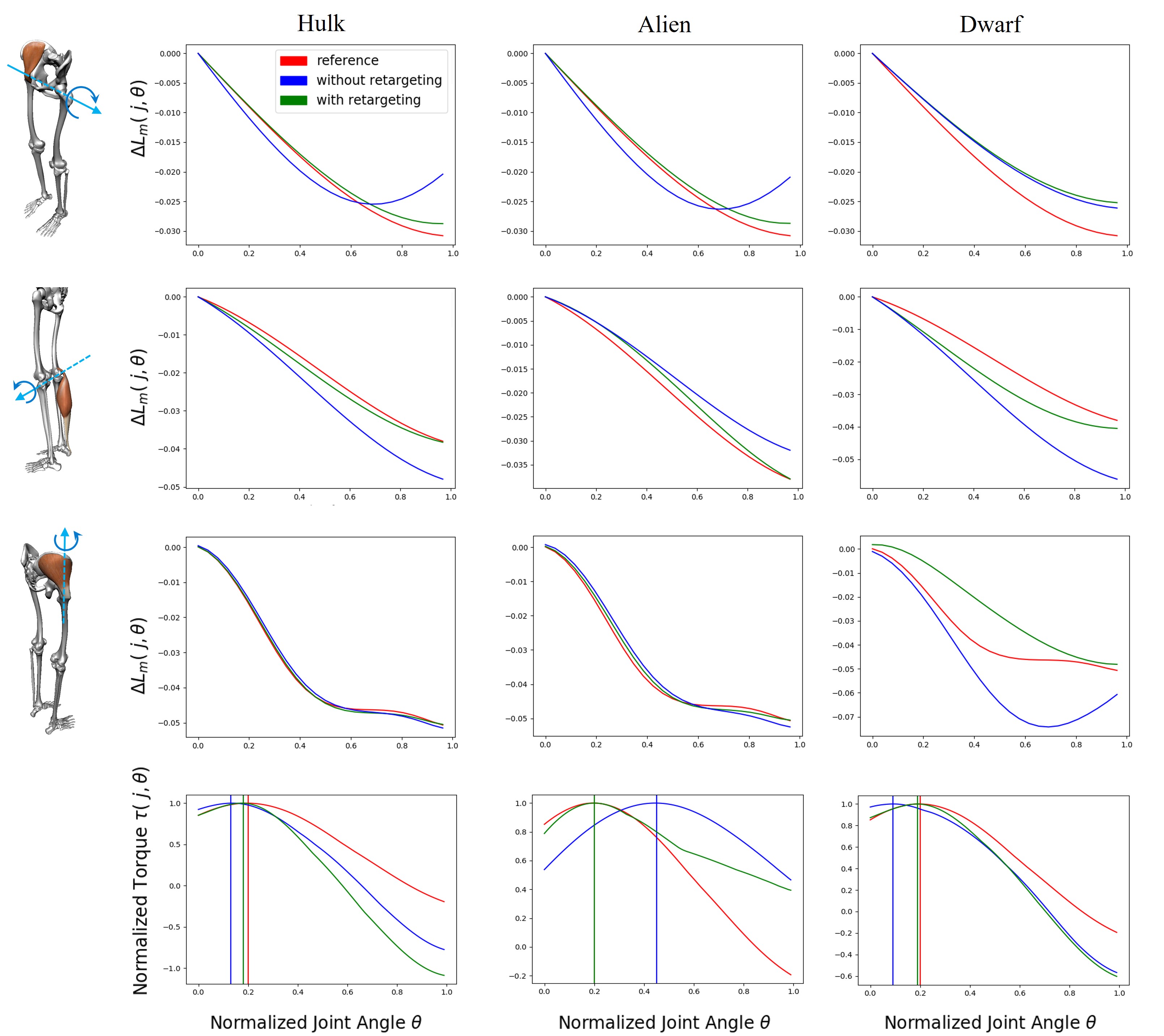}
    \caption{ Musculature retargeting experiments with Hulk, Alien, and Dwarf characters. Length-angle curves (top three rows) of the reference model and the varied models with/without retargeting are plotted for comparison. (Top) Iliacus during hip flexion. (Second row) Gastrocnemius during knee flexion. (Third row) Gluteus Medius during hip medial rotation. (Bottom) The torque-angle curve of the knee shows the angle of peak torque. }
    \label{fig:length_angle_comparison}
\end{center}
\end{figure*}

We found that LBS-based waypoints work noticeably better than fixed-anchor type waypoints in the previous studies~\cite{geijtenbeek2013flexible,lee2014locomotion}. Nonetheless, transplanting the musculature of the reference model to varied skeletons without retargeting often results in undesirable distortion in length-angle curves and disturbance in muscle force directions. Figure~\ref{fig:length_angle_comparison} shows the length-angle curves of the reference and varied models with and without retargeting. The qualitative and quantitative changes of muscle curves can be observed in many major muscles (e.g., Iliacus for the Hulk and the Alien, and Gluteus Medius of the Dwarf in the lower limbs).
The influence on force direction is broader. The force direction at almost all muscle origins and insertions are disturbed by the change of body proportions(see Figure~\ref{fig:force_direction}). Both curve distortions and force direction disturbances affect the maneuverability and ROM of joints. Our muscle routing optimization addresses both problems simultaneously to build a retargeted model that functions as closely as possible to the reference model.

\begin{figure}
\begin{center}
    \includegraphics[width=\columnwidth]{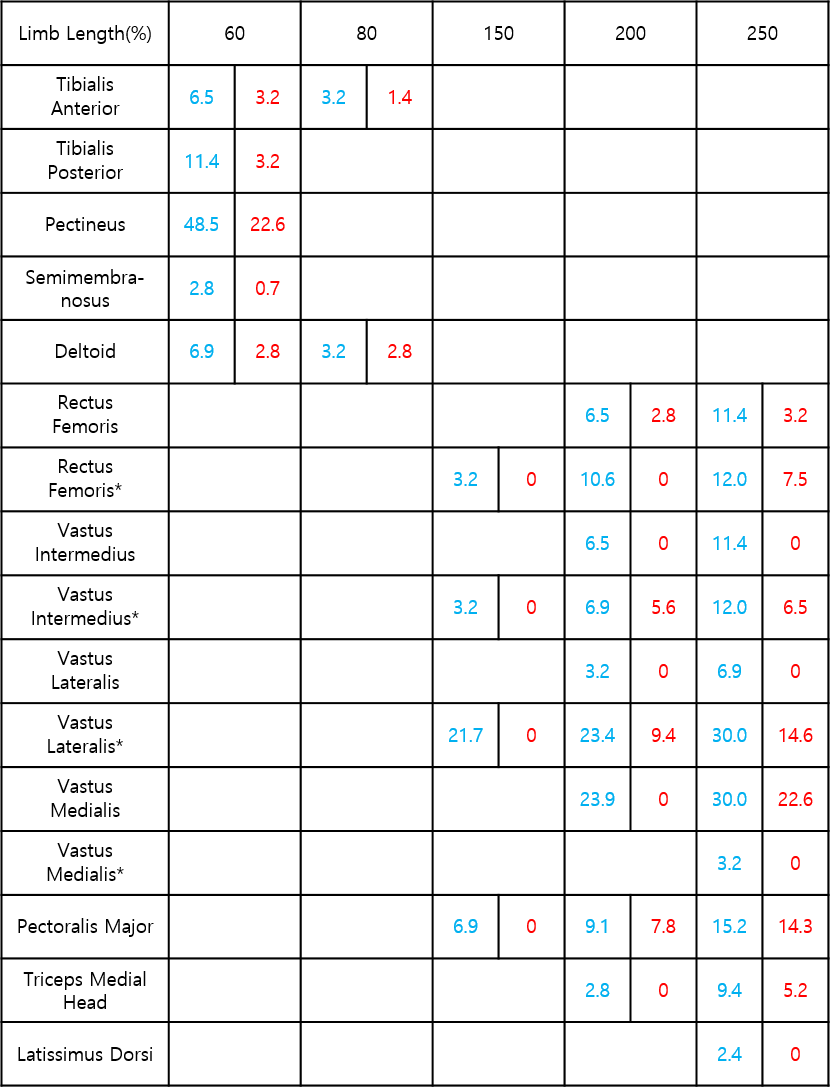}
    \caption{ The degree of functional disorders for the change of limb lengths with (red) and without (blue) retargeting
    }

    \label{fig:func_leg}
\end{center}
\end{figure}

Figure~\ref{fig:func_leg} shows that many muscles change their functional role as the limbs lengthen or shorten in the range of 60\% to 250\% of their original length. Each value indicates the percentage of functional disorders (increasing-to-decreasing and vice versa) in the length-angle curve during joint motion with and without retargeting. Some muscles (e.g., Tibialis Anterior, Tibialis Posterior, Pectineus, and Semimembranosus) change their function with shorter limbs, while functional disorders occurs for some muscles (e.g., Rectus Femoris and Vastus muscles) with longer limbs. For example, the functional disorders for Pectineus muscles occur with shorter (60\% of its original length) legs in half of the range of hip extension and flexion. Our retargeting algorithm alleviates this issue substantially with mild body shape variations, though the problem is not completely fixed with extreme body proportions (e.g., longer limbs by 250\%).

\subsection{Muscle-Coordination using QP}

We tested the usability of our retargeted models in our muscle-actuated simulation system. We briefly summarize the dynamics formulation and its control methods based on Quadratic Programming (QP). The Lagrangian dynamics of the  musculoskeletal model is defined by
\begin{equation}
    \label{eq:lagrangedynamics} \mathbf{M}(\mathbf{q})\ddot{\mathbf{q}} + \mathbf{c}(\mathbf{q},\dot{\mathbf{q}}) = \sum_{m}\mathbf{J}_m^\top\mathbf{f}_m(a_m) + \mathbf{J}_{\textrm c}^\top\mathbf{f}_{\textrm c} + \mathbf{\tau}_{\textrm {ext}}
\end{equation}
where $\mathbf{q}$, $\mathbf{M}(\mathbf{q})$, and $\mathbf{c}(\mathbf{q},\dot{\mathbf{q}})$ are generalized coordinates, the mass matrix, and Coriolis/centrifugal force, respectively. $\mathbf{f}_m$ and $\mathbf{f}_{\textrm c}$ are muscle and constraint forces, respectively. Jacobian matrices $\mathbf{J}_m$ and $\mathbf{J}_{\textrm c}$ map the forces to generalized coordinates. $\tau_{\textrm {ext}}$ is external forces. We compute the constraint forces by solving linear complementary conditions:
\begin{equation}
    \mathbf{v}_{\textrm c} \geq 0, \;\;\;
    \mathbf{f}_{\textrm c} \geq 0, \;\;\;\text{and}\;\;\;
    \mathbf{v}_{\textrm c}^{\top} \mathbf{f}_{\textrm c} = 0,
\end{equation}
where $\mathbf{v}_{\textrm c}=\mathbf{J}_{\textrm c}\dot{\mathbf{q}}$ is the constraint velocity representing the rate of change of the constraint.

Exercising a particular joint is an under-specified control problem, since there are more muscles crossing the joint than minimally required to actuate the joint. Solving the control problem requires muscle coordination to decide activation levels at the muscles that achieve desired joint accelerations $\ddot{\mathbf{q}}_{\textrm d}$. A typical solution method based on QP formulates the problem such that
\begin{equation}
\label{eq:QP}
\begin{aligned}
& \underset{\mathbf{a}}{\textrm{min}}
& & \Vert \ddot{\mathbf{q}}_{\textrm d} - \ddot{\mathbf{q}}(\mathbf{a})  \Vert ^2 + w_{\textrm{reg}} \Vert \mathbf{a} \Vert ^2 \\
& \text{subject to}
& &  
\mathbf{M}\ddot{\mathbf{q}} + \mathbf{c} = \sum_{m}\mathbf{J}_m^\top\mathbf{f}_m(a_m) + \mathbf{J}_{\textrm c}^\top\mathbf{f}_{\textrm c} + \mathbf{\tau}_{\textrm {ext}}\\
&&& 0\leq a_i\leq1 \;\;\; {\textrm {for}} \;\;\;\forall i,
\end{aligned}
\end{equation}
where $\mathbf{a}$ is a vector concatenating muscle activations. This formulation solves for muscle activations while satisfying the Lagrangian equation of motion and regularizing large activations.

\begin{figure}
\begin{center}
    \includegraphics[width=\columnwidth]{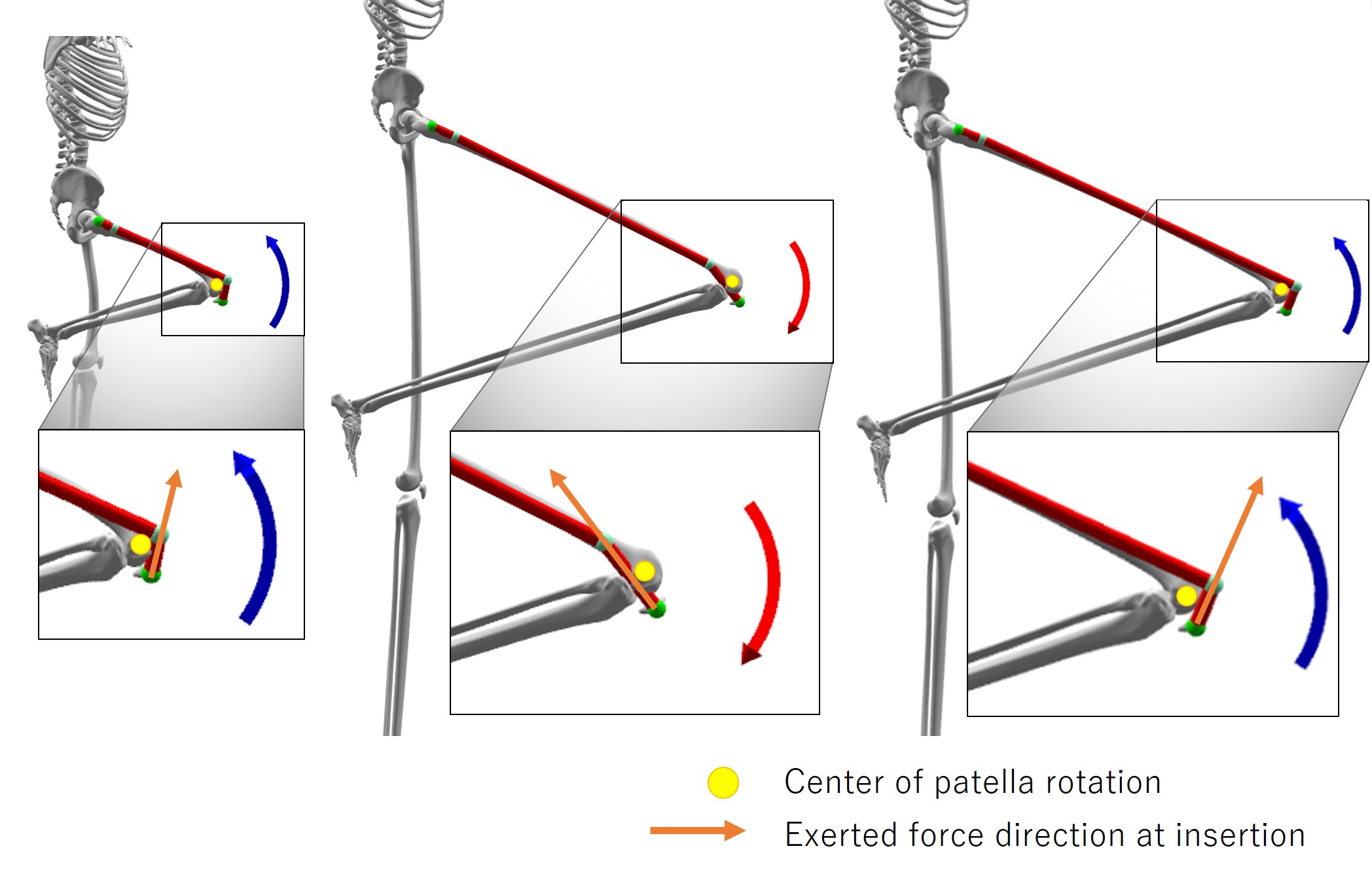}
    \caption{ Vastus lateralis is supposed to be a knee extensor. Its function may be disoriented if skeleton scaling disturbs the muscle force direction at its insertion. (Left) The vastus lateralis acts as expected in the reference model. (Middle) Na\"ive scaling of the musculoskeleton results in its opposite action. (Right) Our retargeting algorithm recovers the force direction and its action.}
    \label{fig:force_direction}
\end{center}
\end{figure}

The formulation allows us to move the joint in an arbitrary direction by specifying desired joint accelerations if the musculoskeletal model is carefully calibrated. The QP with a varied model without retargeting often fails to solve for muscle activations because force directions are degenerate and unable to span the solution space. Our retargeting algorithm recovers the force directions to better condition the problem (see Figure~\ref{fig:force_direction}). The results may be best viewed in the supplementary video.

\begin{figure*}
\begin{center}
    \includegraphics[width=\textwidth]{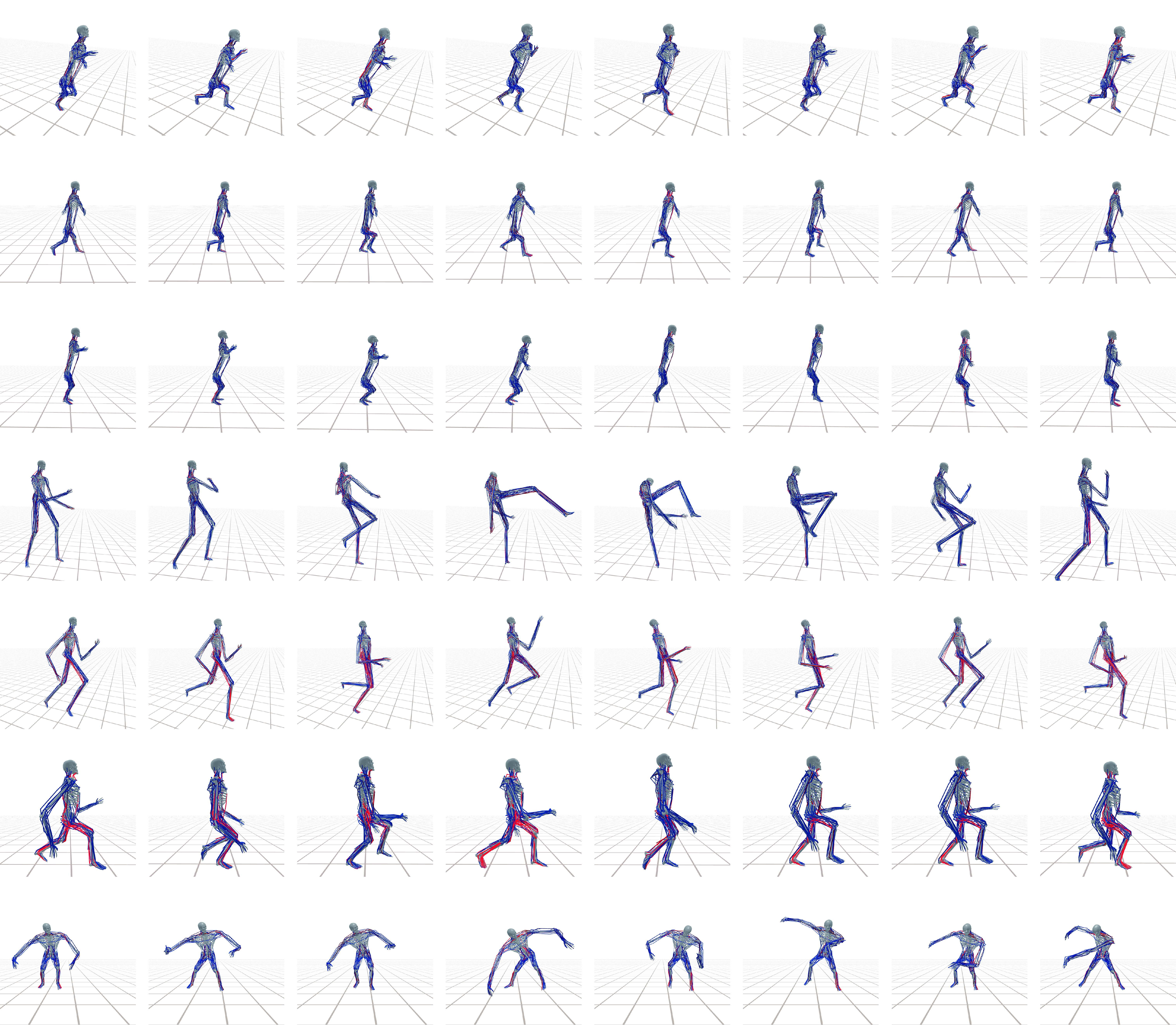}
    \caption{ Our characters learned a variety of motor skills using DRL. All results are physically simulated and controlled by modulating the level of muscle activations and solving muscle contraction dynamics. }
    \label{fig:DRL_results}
\end{center}
\end{figure*}

\subsection{Learning motor skills using DRL}

Our characters also learned to perform challenging full-body motor skills. Deep Reinforcement Learning has recently shown its promise in the control of physically-simulated bipeds and muscle-actuated skeletal figures. The goal of trajectory tracking control is to learn a control policy that imitates reference motion data in physics simulation. It has been shown in previous studies that the control policy represented by deep networks can reproduce highly-detailed human movements~\cite{peng2018deepmimic,lee2019scalable}. 

Let $\mathbf{s}_t$ be the dynamic state of all musculotendon units at time $t$. The state includes muscle lengths and the rate of their length changes. Let $\mathbf{a}_t$ be the level of muscle activations and $r_t$ be a reward of taking action $\mathbf{a}_t$ at state $\mathbf{s}_t$. The agent interacts with the environment by taking actions according to its control policy $\pi_\theta$. The objective of policy learning is to find a control policy that maximizes the expected cumulative rewards
\begin{equation}
\theta^* = \operatorname*{arg\,max}_{\theta} \mathbb{E}_{\mathbf{s}_0,\mathbf{a}_0,\mathbf{s}_1,\cdots} \biggl[ \sum_{t=0} \gamma^t r_t\biggr],
\end{equation}
where $\gamma$ is a discount factor.


The reward evaluates how well joint angles and end-effector positions match the reference trajectory. The reward $r = r_q r_e$ is defined by a multiplication of a joint angle match reward and an end-effector match reward, where
\begin{equation}
\begin{split}
    r_q & = \exp\big(-\sigma_q \sum_j \|\hat{\mathbf{q}}_j(t) \ominus \mathbf{q}_j(t)\|^2\big) \\
    r_e & = \exp\big(-\sigma_e \sum_e \|\hat{\mathbf{p}}_e(t) - \mathbf{p}_e(t)\|^2\big),
\end{split}
\end{equation}
where $\mathbf{q}_j$ are the joint configurations in generalized coordinates and $\mathbf{p}_e$ are end-effectors positions. $j$ and $e$ are the indices of joints and end-effectors, respectively. Our implementation of the end-effector reward includes the positions of both hands, both feet, and the head relative to the moving coordinate frame of the skeletal root (pelvis). The hat symbol indicates the positions and joint configurations taken from the reference trajectory. The joint configurations are represented by unit quaternions. The geodesic distance $\mathbf{q}_1\ominus\mathbf{q}_2=\ln(\mathbf{q}_2^{-1}\mathbf{q}_1)$ measures the shortest rotation angle between two joint configurations~\cite{lee2008geometric}. We employed a hierarchical RL algorithm proposed by Lee et al.~\shortcite{lee2019scalable} to learn muscle-actuated control policies mimicking reference motion data.

We collected motion capture data available on the web and retargeted each motion clip to our kinematic models using {\em Autodesk MotionBuilder$^\mathrm{TM}$}. The DRL algorithm is able to learn muscle-actuated control policies successfully with mild body shape variations.  The DRL algorithm can also cope with larger variations to a certain extent. The highly-varied characters were able to successfully imitate motor skills, including walking, running, jumping, kicking, and dancing while maintaining their balance (see Figure~\ref{fig:DRL_results} and the results may be best viewed in the accompanied video).

\subsection{Moment Arm}

The moment arm of a muscle $m$ to a joint $j$ is a quantity of interest often referred to by biomechanics literature. The moment arm is the perpendicular distance from the joint to the line of muscle tension force and serves as a measure of effectiveness that the muscle has on joint torque. The moment arm about a particular axis $\hat{n}$ of action at joint angle $\theta$ can be calculated by~\cite{lieber1993skeletal}
\begin{equation}
    r_{\theta} = \frac{\tau_\theta}{f}
    = \frac{J_m^\top \mathbf{f}_m (a_m) \cdot \hat{n}}{f(a_m)}
\end{equation}
where $\tau_\theta$ is the effective torque applied to joint $j$ by muscle $m$ and $f(a_m)$ is the tension force at the attachment site generated by muscle activation.

\begin{figure*}
\begin{center}
    \includegraphics[width=.88\textwidth]{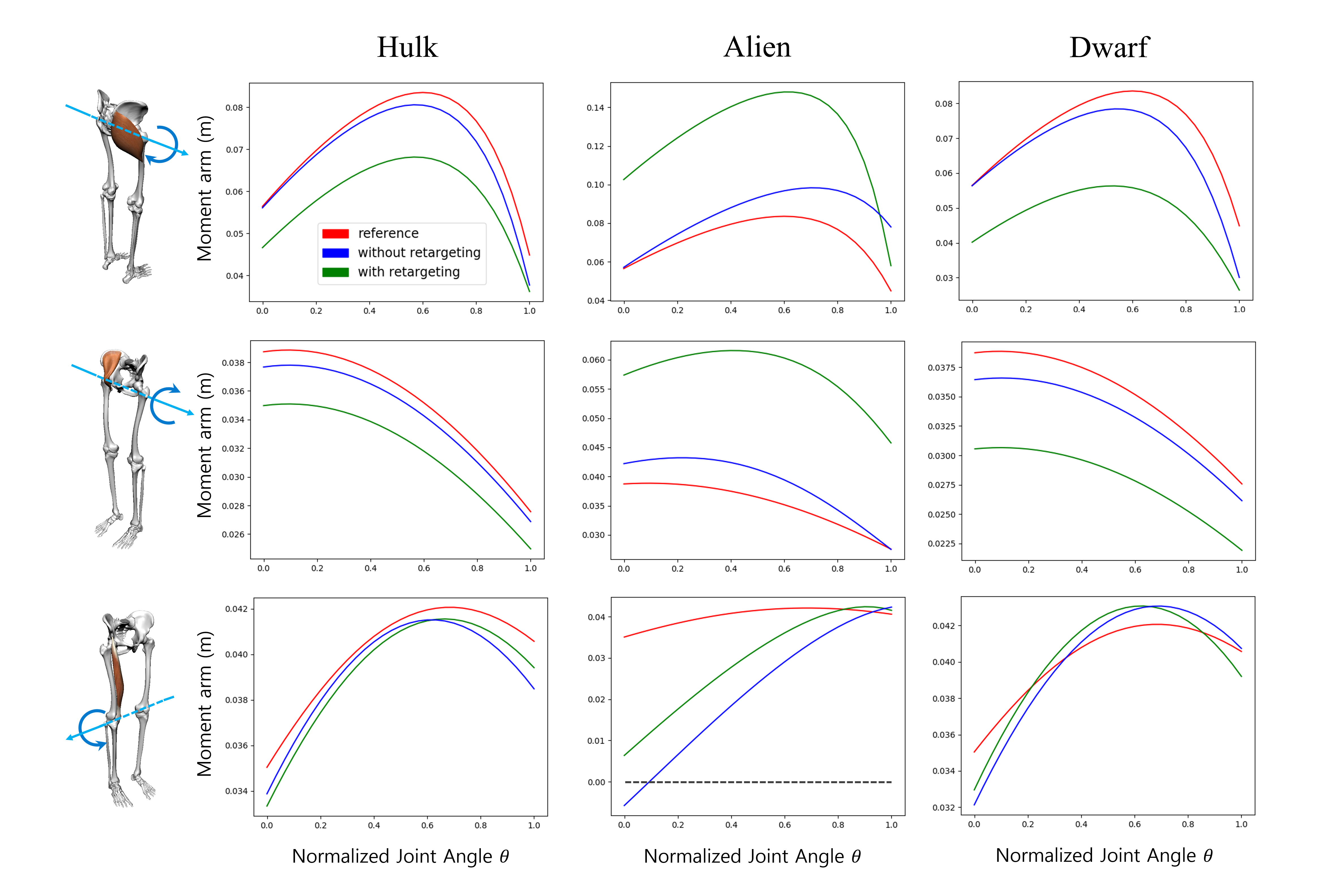}
    \caption{ The relation between moment arm and joint angle. (Top) Gluteus Maximus during hip extension. (Middle) Iliacus during hip flexion. (Bottom) Rectus Femoris during knee extension.}
    \label{fig:ma_angle_comparison}
\end{center}
\end{figure*}


Moment arm is not an invariant factor we want to preserve across different body sizes and proportions. We rather wish to have moment arms to be adjusted appropriately during the retargeting process. In Figure~\ref{fig:ma_angle_comparison}, the angle-moment arm curves in the retargeted models differ from the curves in the reference model. The Alien has larger moment arms for hip flexors and extensors to compensate for longer Femurs. As the femurs get longer, larger moment arms are preferred to use flexor and extensor muscles more efficiently. The retargeting algorithm reduces the moment arms for the Hulk and the Dwarf, on the contrary, to compensate for their short Femurs.



\subsection{Individualized Anatomy Modeling}

We would like to create individualized anatomic models from medical images and physical examination data. There are standard physical examination procedures of measuring joint ROMs in physical practice~\cite{thomas1974diseases, silfverskiold1924reduction, staheli1977prone, lee2011reliability}, such as popliteal angle, Staheli, and  Silfverski\"old tests. In this example, we used biplanar X-ray images of an anonymous person to create a skeletal model and demonstrated how ROM editing can be performed according to physical examination data.

\begin{figure}
\begin{center}
    \includegraphics[width=.8\columnwidth]{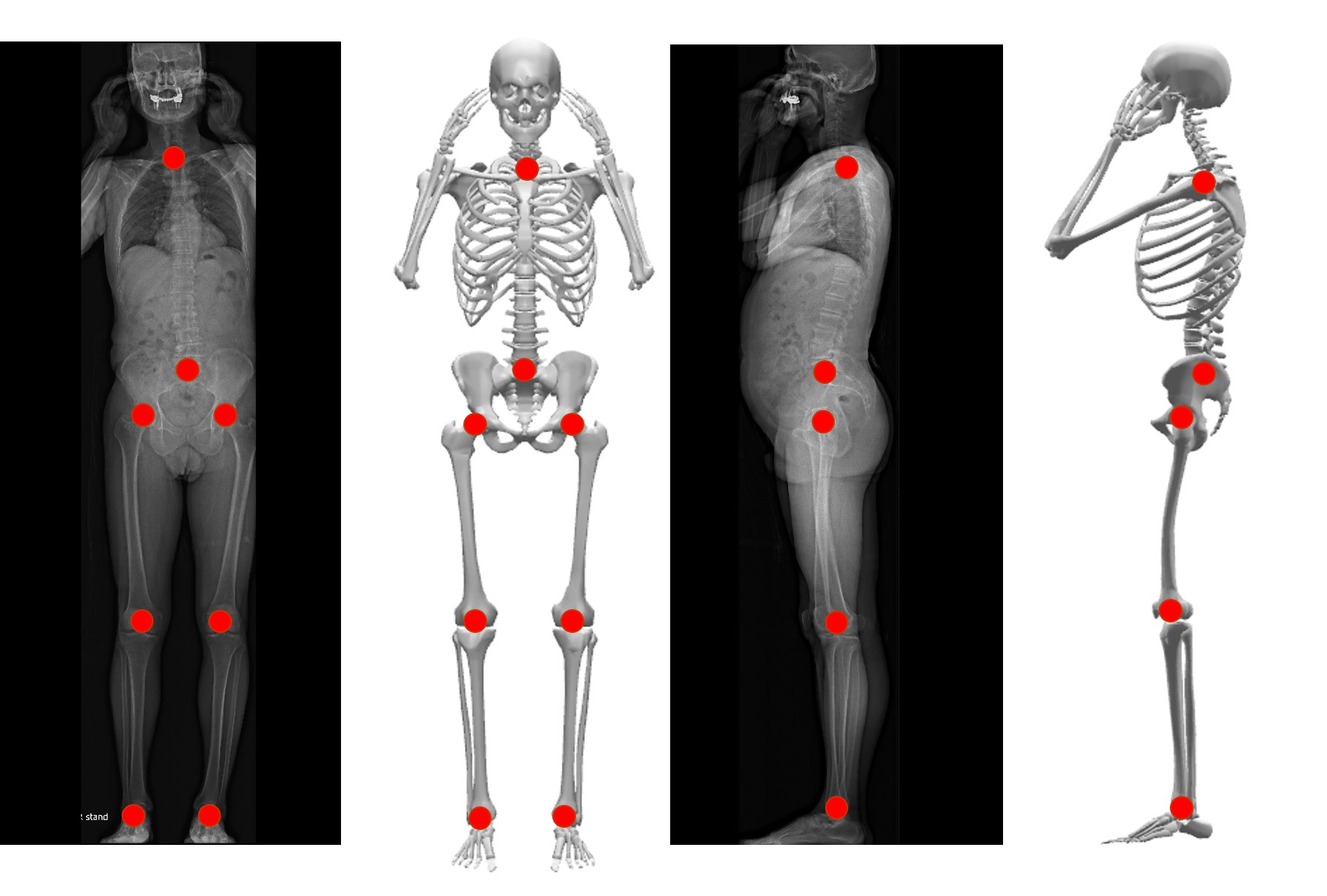}
    \caption{ The 3D skeleton reconstructed from biplanar X-ray images.}
    \label{fig:biplanar}
\end{center}
\end{figure}

The biplanar image data include lateral and anteroposterior X-ray images acquired simultaneously at precisely calibrated view angles (see Figure~\ref{fig:biplanar}). The calibrated orthographic images allow us to reconstruct 3D points from a pair of 2D feature points on the images. We selected joint points on the X-ray images and scaled our parametric skeleton model to match the reconstructed 3D joint points. Musculature retargeting to the scaled skeleton creates an individualized musculoskeletal model. 

\begin{figure}

\begin{center}
    \includegraphics[width=\columnwidth]{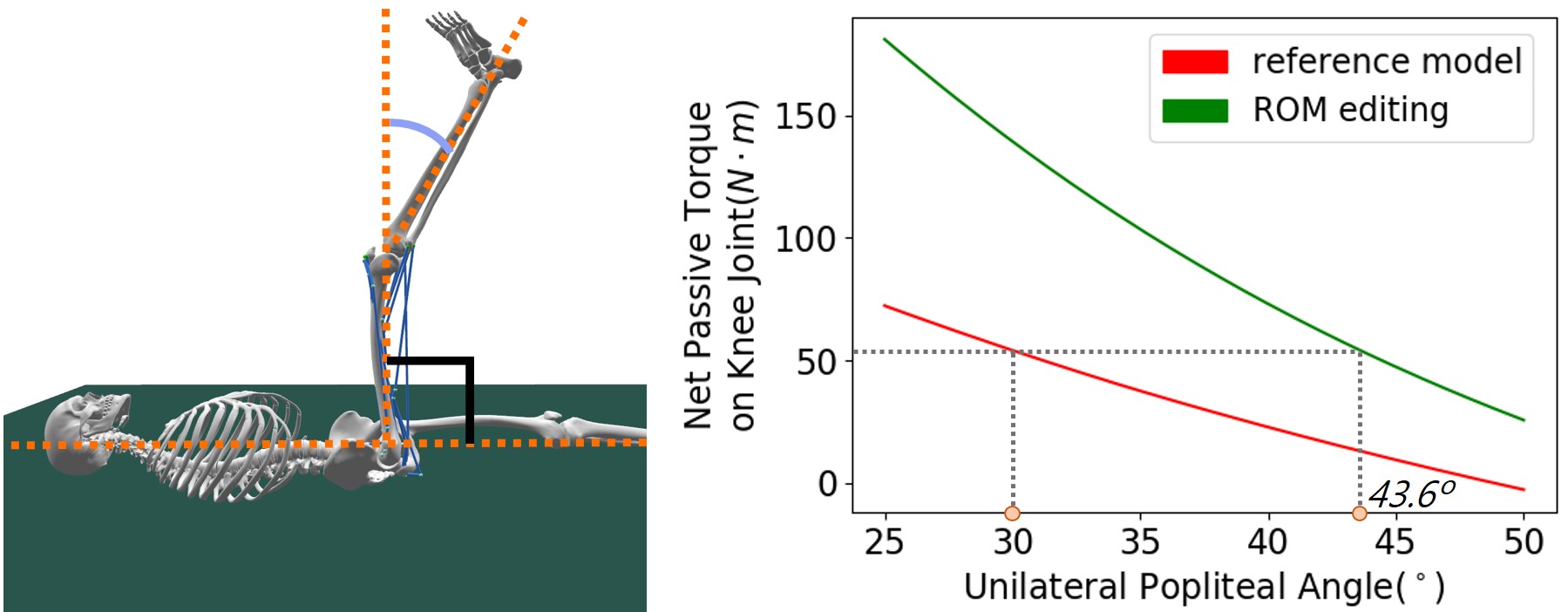}
    \caption{ The unilateral popliteal angle of our reference model is 30$^\circ$ assuming that the threshold for the net passive torque is 50 $N\cdot m$. ROM editing can enhance or reduce the ROM in the knee. The reduced ROM shown in the green curve results in a larger unilateral popliteal angle.
    }
    \label{fig:popliteal_angle}
\end{center}
\end{figure}

The ROM measurements in physical examination data can be incorporated into our anatomic model. For example, the unilateral popliteal angle is a test for measuring hamstring tightness, which is related to the ROM in the knee. The test measures the range of a knee angle when the subject is in the supine position with its leg flexed to 90$^\circ$ at the hip and the knee is extended passively. We simulated this procedure to have our musculoskeletal model take the specified pose (see Figure~\ref{fig:popliteal_angle}). Interactive ROM editing allows us to change the knee ROM to match any target range by deciding the scaling and shifting factors as explained in Section~\ref{sec:rom_editing}.









\section{Discussion}

Our attempt to build a musculoskeletal model for a variety of body shapes is an important step toward the goal of expanding the applicability of anatomic human modeling and simulation. Nonetheless, our current framework still has several limitations in its scope of applicability and biomechanical accuracy. Human anatomy is an extremely complicated system. There has been a trail of research works that incrementally added details to the computer model of human anatomy. Currently, our model ignores a lot of important anatomic features and phenomena, such as ligaments and contact coupling. Joint ROMs are largely influenced by the collision between bony features and tensions generated by ligaments around the joints. Muscle paths wrapping around bones and other muscles will best be approximated by contact coupling between anatomic features if their high-resolution geometry and volumetric FEM simulation are provided~\cite{lee2018dexterous}. The use of waypoints is a crude, yet computationally-feasible approximation.




Our demonstration of reconstructing a musculoskeletal model from biplanar X-rays and clinical examination shows the potential applicability of our study in clinical practice.
Ideally, we wish to be able to construct individualized musculoskeletal models for any human subjects with minimal effort. Although the potential is apparent, we have not tried to validate or verify its effectiveness in medical applications, which may require rigorous experiments with human subjects~\cite{lee2015push}. This ambitious goal remains the subject of future research beyond the scope of this paper.

Our algorithm assumes the presence of a reference model, which can be retargeted only to structurally equivalent bodies. If we want to build a musculoskeletal model of non-anthropomorphic characters, there is no high-quality reference model to begin with. An interesting direction for future research is to build anatomical models from scratch, potentially with help of various types of measurements including 3D scans, medical images, EMG, motion capture, and videos~\cite{kadlevcek2016reconstructing}.
\section*{Acknowledgements}

This work was supported by Samsung Research Funding Center under Project Number SRFC-IT1801-01.

\bibliographystyle{eg-alpha-doi} 
\bibliography{reference}       


\newpage

\end{document}